\documentclass[conference]{IEEEtran}
\IEEEoverridecommandlockouts
\PassOptionsToPackage{table}{xcolor}

\usepackage{multirow}
\usepackage{pifont}
\usepackage{algorithmicx}
\usepackage{algorithm}
\usepackage{algpseudocode}
\usepackage{tcolorbox}

\usepackage{amsmath}

\usepackage{pgfplots}
\pgfplotsset{compat=newest}
\usepackage{pgfplotstable}
\usepgfplotslibrary{groupplots}
\usepackage{tikz}
\usepackage{listings}
\usepackage{textcomp}
\usepackage{graphicx} 
\usepackage{multicol}
\usepackage{subfigure}
\usepackage{epsfig}

\usepackage{svg}

\usepackage{booktabs}
\usepackage{array}

\usepackage[table]{xcolor} 
\usepackage{colortbl} 

\begin{document}

\title{Building an OceanBase-based Distributed Nearly Real-time Analytical Processing Database System\\
}


\author{\IEEEauthorblockN{Quanqing Xu, Chuanhui Yang\IEEEauthorrefmark{1}\thanks{*Chuanhui Yang is the corresponding author.}, Ruijie Li, Dongdong Xie, Hui Cao, 
Yi Xiao, Junquan Chen, \\
Yanzuo Wang, Saitong Zhao, Fusheng Han, Bin Liu, Guoping Wang, Yuzhong Zhao, Mingqiang Zhuang}
\IEEEauthorblockA{\textit{OceanBase, Ant Group} \\
OceanBaseLabs@service.oceanbase.com}
}

\maketitle

\begin{abstract}
The growing demand for database systems capable of efficiently managing massive datasets while delivering real-time transaction processing and advanced analytical capabilities has become critical in modern data infrastructure. While traditional OLAP systems often fail to meet these dual requirements, emerging real-time analytical processing systems still face persistent challenges, such as excessive data redundancy, complex cross-system synchronization, and suboptimal temporal efficiency. This paper introduces OceanBase Mercury as an innovative OLAP system designed for petabyte-scale data. The system features a distributed, multi-tenant architecture that ensures essential enterprise-grade requirements, including continuous availability and elastic scalability. Our technical contributions include three key components: (1) an adaptive columnar storage format with hybrid data layout optimization, (2) a differential refresh mechanism for materialized views with temporal consistency guarantees, and (3) a polymorphic vectorization engine supporting three distinct data formats. Empirical evaluations under real-world workloads demonstrate that OceanBase Mercury outperforms specialized OLAP engines by $1.3\times$ to $3.1\times$ speedup in query latency while maintaining sub-second latency, positioning it as a groundbreaking AP solution that effectively balances analytical depth with operational agility in big data environments.
\end{abstract}


\section{Introduction}

In the rapidly evolving landscape of big data, enterprises increasingly require database systems that seamlessly handle vast volumes of data while simultaneously providing nearly real-time transactional processing and robust analytical capabilities. 
Service workloads necessitate the execution of intricate queries on substantial volumes of newly ingested data, with stringent requirements for sub-second latency.
This requires systems to efficiently process data with guaranteed freshness, deliver rapid responses, and maintain high performance as data is continuously updated.
Traditional Online Analytical Processing (OLAP) systems often fall short in addressing these dual demands, as they typically conduct complex data analysis on relatively static data with coarse time granularity for data freshness. Some existing big data solutions adopt a combination of different systems to address hybrid serving and analytical processing, resulting in unsatisfactory time and storage efficiency due to excessive data duplication and complex data synchronization across systems.

The importance of hybrid databases is underscored by recent innovations in the field. For example, HyPer~\cite{DBLP:conf/birte/KemperN10} uses hardware-assisted replication to maintain consistent snapshots of transactional data, enabling concurrent HTAP (Hybrid Transaction/Analytical Processing) operations without significant performance degradation. PolarDB~\cite{DBLP:conf/fast/CaoLCZLWOWWKLZZ20} integrates computational storage to optimize analytical performance in a cloud environment, showcasing how cloud-native databases can efficiently support analytical workloads. Greenplum~\cite{DBLP:conf/sigmod/LyuZXGWCPYGWLAY21} extends traditional OLAP capabilities by incorporating Online Transaction Processing (OLTP) features, utilizing a Massively Parallel Processing (MPP) architecture to manage and analyze petabytes of data across distributed nodes. AnalyticDB-V~\cite{DBLP:journals/pvldb/WeiWWLZ0C20}, with its query fusion capabilities for structured and unstructured data, represents a sophisticated approach to hybrid analytical engines.

Moreover, recent market solutions like YouTube's Procella~\cite{DBLP:journals/pvldb/ChattopadhyayDL19}, Alibaba's Hologres~\cite{DBLP:journals/pvldb/JiangHXJJXJYWJM20}, and ByteDance's Krypton~\cite{DBLP:journals/pvldb/ChenSCZLDFWXCDG23} further explore the systems that unify serving and analytical data processing. Procella reduces system complexity while enhancing performance by unifying the two workloads within a single framework. Similarly, Hologres demonstrates the advantages of combining transactional and analytical processing capabilities within a cloud-native service. Krypton exemplifies the balance between near real-time serving and analytical workloads, leveraging an integrated SQL engine to handle diverse data processing needs efficiently. 
Although these systems have effectively improved data-freshness performance and data analysis efficiency, developing a competitive OLAP system with novel technologies and features remains a challenging and exciting exploration. However, existing HTAP systems face trade-offs: some require separate analytical infrastructure (increasing complexity), while others struggle with seamless column store integration that maintains transactional capabilities. OceanBase Mercury's hybrid design (columnar baseline, row-based incremental) uniquely enables full DML operations with analytical performance, achieving configurable freshness (from zero to milliseconds) without separate systems.


OceanBase demonstrates web-scale capabilities through its proven deployment handling trillion-row datasets ($\approx$20PB) across 1,500-node clusters~\cite{OB_github,DBLP:journals/pvldb/YangYHZYYCZSXYL22}.
This paper proposes the OceanBase OLAP database system named Mercury\footnote{Mercury has incredible speed in Roman religion and mythology, so it is used to name our OLAP database system. It is an alias for OceanBase 4.3.3.}, designed to operate at petabyte scale~\cite{DBLP:journals/pvldb/YangYHZYYCZSXYL22, TPCC-OceanBase} with near real-time analytical processing capabilities, e.g., online review service and high-definition
map in AMap~\cite{xu2025oceanbase}. OceanBase Mercury builds upon these innovations, offering a distributed, multi-tenant architecture that ensures high availability, scalability, and robust performance. Its design addresses the unique challenges of maintaining ACID (Atomicity, Consistency, Isolation, Durability) properties across distributed transactions~\cite{zhang2023efficient} while enabling complex analytical queries. By integrating both OLTP and OLAP functionalities within a single system, it supports near real-time business intelligence and operational analytics, crucial for modern enterprise applications. It employs a combination of techniques such as multi-version concurrency control (MVCC) for consistency, partition-based distribution for scalability, and sophisticated replication mechanisms to ensure high availability and fault tolerance.







Therefore, the contributions of this paper are as follows:
\begin{itemize}
    \item \textbf{Novel Hybrid Storage Architecture.} We propose an adaptive column store within the LSM-tree architecture. By maintaining incremental data in row format for transactional efficiency and baseline data in columnar format for analytical speed, our design enables full DML capabilities on row stores without the update penalties typical of traditional OLAP systems.

    \item \textbf{Efficient Materialized Views.} We optimize the implementation of materialized views (MV) to ensure data consistency and query efficiency. Additionally, we propose efficient refresh mechanisms for materialized views, including full refreshes using hidden tables and incremental refreshes using materialized view logs. 

    \item \textbf{Optimized Vectorized Engine.} We design a vectorized execution engine tailored for this hybrid layout. We specially designed three data formats to adapt to various computing scenarios. We redesigned and optimized several operators and expressions for storage vectorization, improving the execution efficiency of vectorized engines.

    \item \textbf{Deployment and Evaluation.} We have deployed Mercury in OceanBase's internal big data stack as well as in public cloud offerings and conducted a thorough performance study under various workloads. Our results show that OceanBase Mercury achieves superior performance even compared with specialized OLAP engines.
\end{itemize}

The paper is organized as follows. \S\ref{sec:architecture} provides an overview of the system architecture of OceanBase Mercury. \S\ref{sec:ColumnStore} describes the column store in OceanBase Mercury. In \S\ref{sec:MaterializedView}, we discuss the materialized view, and in \S\ref{sec:VectorizedExecutionEngine}, we present a vectorized execution engine for batch processing of data. We present the experiments in \S\ref{sec:evaluation} to prove the high availability and high performance of OceanBase Mercury. We review the related work in \S\ref{sec:related} and discuss the lessons and future innovations in \S\ref{sec:lesson}. Finally, we give the conclusions in \S\ref{sec:conclusion}. OceanBase is an open-source project under Mulan Public License 2.0~\cite{MulanPubL-2.0} and the source code is available on both Gitee~\cite{OB_gitee} and GitHub~\cite{OB_github}.

\section{System Architecture}
\label{sec:architecture}

\subsection{OceanBase Overview}

OceanBase~\cite{DBLP:journals/pvldb/YangYHZYYCZSXYL22,DBLP:journals/pvldb/YangXGYWZKLWX23} uses a shared-nothing architecture to build a distributed cluster. Each node in the cluster is equipped with an independent SQL engine, transaction engine, and storage engine, ensuring excellent scalability, high availability, high performance, low cost, and high compatibility with mainstream databases. The multiple nodes in the cluster are aggregated into several availability zones. 
Each zone contains multiple processes called OBServers that process data and are marked with two attributes: Internet Data Center (IDC) and the region or city the IDC locates, providing geographic information for disaster recovery and high availability design. In response to different needs, OceanBase provides a variety of high-availability deployment strategies, including three data centers and three replicas in the same city, five centers and five replicas in three locations, primary-secondary deployment modes in the same city and different cities, and multi-IDC deployment solutions combined with arbitration services, making full use of the geographical distribution characteristics of each zone to enhance the resilience and disaster recovery capabilities of the system. For further details on OceanBase’s architecture design, readers may refer to ~\cite{DBLP:journals/pvldb/YangXGYWZKLWX23}.

\subsection{System Architecture}

\begin{figure}[htp]
\vspace{-6pt}
    \centering
    \includegraphics[width=0.9\linewidth]{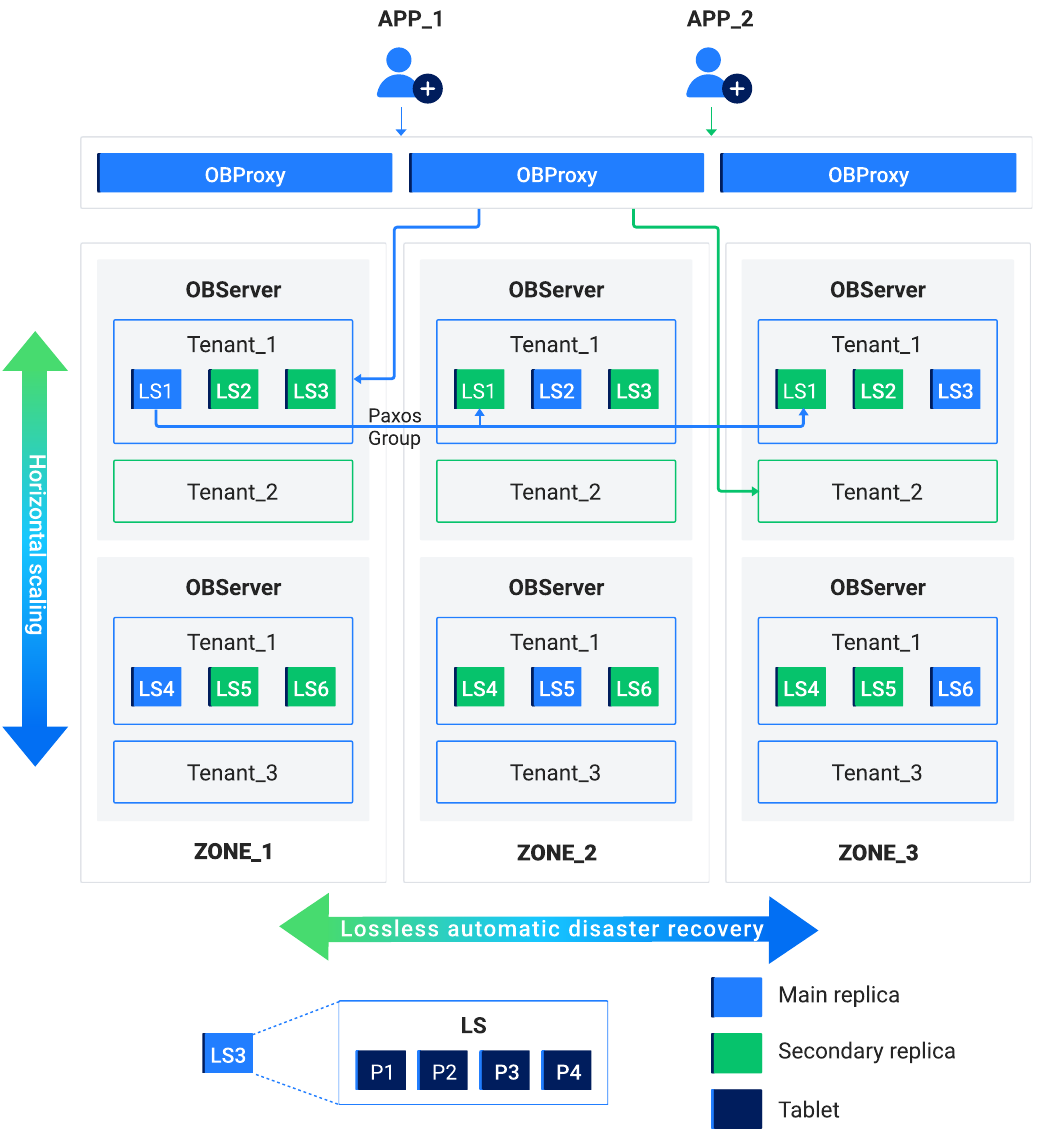}
    \vspace{-6pt}
    \caption{System Architecture}
    \label{fig:SystemArchitecture}
    \vspace{-3pt}
\end{figure}

As illustrated in Figure~\ref{fig:SystemArchitecture}, OceanBase implements horizontal sharding for single-table data through user-defined partitioning rules, enabling distributed storage of table data across multiple partitions. This architecture significantly enhances data management flexibility and processing efficiency. Building upon this foundation, the system employs a two-tiered partitioning mechanism: primary partitions serve as logical groupings for data organization, while secondary partitions represent physical implementations. Each physical partition is mapped to a tablet --- a fundamental storage unit that sequentially orders data records. This hierarchical design optimizes storage efficiency by aligning logical and physical structures, simultaneously enhancing data access performance and enabling horizontal scalability through distributed tablet management.

As shown in Figure~\ref{fig:SystemArchitecture}, OceanBase ensures data persistence through an immediate Redo log recording mechanism. Whenever a user modifies tablet data, the system writes these changes to the Redo log stream specifically associated with that tablet. Both log streams~\cite{HanXuVLDB24} and tablets employ a multi-replica architecture, with replicas strategically distributed across different availability zones to mitigate single-region disaster risks. In this design, one primary replica (Leader) handles all write operations, while secondary replicas (Followers) maintain synchronization and consistency through the Multi-Paxos consensus protocol. Upon primary replica failure, a quorum-based election process enables surviving replicas to rapidly establish a new Leader, guaranteeing uninterrupted service continuity through automated failover mechanisms.

Each cluster node runs an OceanBase server process (OBServer), implemented as a multi-threaded architecture. This process performs two primary functions: 1) local data management by accessing partitioned data stored on the host node, and 2) SQL execution by parsing and processing queries routed to the machine. Inter-node communication occurs through TCP/IP protocol, enabling reliable data exchange between distributed components. It also acts as a client interface endpoint, listening for incoming connection requests from external applications, establishing database sessions, and delivering transactional services through standardized SQL interfaces.

\subsection{Analytical Processing Engine}

OceanBase Mercury's analytical processing engine is integrated into the existing OBServer architecture, sharing the same storage and replication infrastructure with OLTP workloads. This design choice imposes several architectural constraints and design decisions:

\textbf{Isolation from OLTP Workloads:} The analytical engine operates on column replicas, ensuring that analytical queries do not interfere with ongoing transactional operations. Analytical queries read from consistent column replicas, which allows concurrent read and write operations without blocking. This column replica-based approach ensures that analytical queries see a consistent view of the data at a specific point in time, while transactional writes continue without interruption. However, this design has implications: (1) column replica creation and maintenance incur overhead, particularly for long-running analytical queries that hold column replicas for extended periods; (2) queries merge data from columnar baseline and row-based incremental storage, introducing small latency (typically milliseconds) even when targeting freshness of zero.

\textbf{Batch-Oriented Processing:} The analytical engine is designed for throughput-oriented batch processing rather than optimizing for individual query latency. The vectorized execution engine (described in Section~\ref{sec:VectorizedExecutionEngine}) processes data in batches, maximizing CPU utilization and memory bandwidth. This design aligns with the analytical workload characteristics, where queries typically scan large volumes of data and benefit from high throughput rather than low latency for individual queries. The batch processing paradigm means that the system prioritizes overall query throughput over individual query response time, which is appropriate for analytical workloads but may not be suitable for interactive queries requiring sub-second response times. 

\textbf{Resource Management:} To prevent analytical queries from impacting OLTP performance, the system employs resource isolation mechanisms. Each tenant can be configured with resource limits (CPU and memory), ensuring that analytical workloads do not consume resources needed for transactional operations. Crucially, the internal scheduler is designed to prioritize OLTP threads, ensuring that transactional latencies remain stable even under concurrent analytical loads. 
Despite these architectural safeguards, resource contention—particularly regarding I/O bandwidth—remains a challenge in production environments. Consequently, our deployment experience suggests that heavy analytical queries are best routed to dedicated resource pools or scheduled during off-peak hours to minimize interference.

\textbf{Storage Integration:} The analytical engine leverages OceanBase's existing column store capabilities (described in Section~\ref{sec:ColumnStore}), which stores baseline data in columnar format while maintaining incremental data in row format. This hybrid approach allows the system to benefit from columnar storage's analytical performance while maintaining compatibility with existing row-based transactional workloads. Analytical queries automatically access the appropriate storage format based on the query pattern and data location. However, this design introduces complexity: queries must merge data from both columnar baseline and row-based incremental storage, requiring careful query planning and execution optimization. 

\textbf{OLTP Performance Preservation:}
We safeguard transactional performance through strict architectural guarantees. Incremental data is maintained in row format, ensuring that DML capabilities and transactional write performance remain identical to those of standard row-store tables. Furthermore, OceanBase protects OLTP workloads through multi-layer isolation: (1) OS-level cgroup isolation to prevent analytical queries from exhausting CPU and memory resources; (2) physical read-write separation by placing column store replicas that serve analytical queries on follower replicas; and (3) automatic query routing that redirects heavy analytical workloads to column store followers, thereby preventing interference with OLTP workloads on primary replicas. Further details on OceanBase's transactional capabilities are provided in~\cite{DBLP:journals/pvldb/YangYHZYYCZSXYL22}.

\textbf{Deployment Insights:} Our production deployments across OceanBase's internal big data stack and public cloud offerings reveal several practical considerations. First, the shared infrastructure design means that capacity planning must account for both OLTP and analytical workloads, requiring careful monitoring and resource allocation. Second, the snapshot-based read model means that analytical queries may see slightly stale data depending on snapshot selection, which is acceptable for most analytical workloads but requires clear communication to users about data freshness guarantees. Third, the batch-oriented processing model means that individual query latency can vary significantly depending on data volume and query complexity, with some queries taking milliseconds to seconds for large-scale analytical workloads. This trade-off between throughput and latency is a fundamental design decision that aligns with the analytical workload characteristics we observe in production.

\section{Column Store}
\label{sec:ColumnStore}

Column store plays a pivotal role in boosting the performance of analytical queries. It is a critical feature that enables OceanBase to effectively support OLAP workloads. This section delves into the unique implementation of OceanBase's column store, highlighting its design and optimizations.

\subsection{Baseline and Incremental Data}

OceanBase realizes the integration of row store and column store. It has used the LSM-tree architecture since its inception, accumulating extensive engineering practical experience with this architecture. Additionally, OceanBase has made many special designs for the column store, such as dividing all user data into two categories: baseline data and incremental data.
In columnar store design, data is typically not updated in place. Instead, it is batched on the client or database side and organized into a columnar format through appending or merging. OceanBase's storage engine employs a multi-level LSM-Tree, which inherently avoids in-place updates, making it naturally suitable for columnar store design.

Our columnar store approach organizes the data into a columnar format during the compaction process of the lowest level of the LSM-Tree (baseline data), while the upper-layer incremental data (MemTable and minor SSTables), logs, transaction modules, and other components remain unchanged. This ensures that the row store retains full DML (Data Manipulation Language) capabilities (such as single-row updates and large transactions~\cite{fang2025malt}) and data synchronization capabilities (downstream CDC, Binlog). The novelty is the LSM-tree-integrated hybrid design: baseline data (major compaction) becomes columnar while incremental writes stay row-based. Unlike traditional column stores that sacrifice DML capabilities, this design preserves full transactional operations (single-row updates, large transactions) while achieving analytical performance, enabling true HTAP without separate systems.

\noindent$\bullet\ $\textbf{Baseline data:} OceanBase leverages its distributed multi-replica architecture to implement a proprietary mechanism called ``Daily Compaction''. This process operates through two distinct phases: first, tenants either schedule periodic compaction cycles or trigger them via administrative commands to designate a global version number. Second, all data replicas across the tenant's distributed storage execute a coordinated Major Compaction to synchronize with this version, resulting in the creation of version-specific baseline data. Crucially, this architecture guarantees identical baseline data across all replicas for the same version, ensuring data consistency and system-wide integrity through deterministic compaction outcomes. To address near-real-time OLAP requirements, the query engine performs an on-the-fly merge of this static baseline data with dynamic, row-based incremental data. This ensures that while the conversion to columnar format occurs daily, data visibility is immediate (achieving data freshness $\approx 0$, as evidenced by ``System A2'' in~\cite{zhang2024hybench}), allowing analytical queries to access recently committed transactions without waiting for the compaction cycle.


\noindent$\bullet\ $\textbf{Incremental data:} All data written after the latest version of the baseline data is considered incremental data. Specifically, incremental data can either be the in-memory data that users have just written to the MemTable or the disk data that has already been dumped into SSTable. MemTable and SSTable are components of OceanBase's storage engine, where MemTable resides in memory, storing dynamic data, and provides read and write operations; SSTable resides on disk, storing static data, and is read-only. 

In large-scale production environments, incremental data represents 1\%--10\% of daily writes. While reading this data ensures nearly real-time visibility, the on-the-fly merge introduces overhead: queries accessing only baseline data are typically 5--10$\times$ faster than those merging substantial incremental data. The daily compaction cycle is critical to periodically convert row-based increments into columnar baseline, restoring analytical performance while minimizing resource contention by scheduling execution during off-peak hours.



\subsection{Adaptive Store}

OceanBase uses the characteristics of the baseline and incremental data to propose a set of column store implementation methods that are transparent to the upper layer:

\noindent$\bullet\ $ Baseline data is stored in column store, and incremental data is kept in row store. All user DML operations are not affected.  A column store table can still perform all transaction operations like row store tables.

\begin{figure}[htp]
\vspace{-6pt}
    \centering
    \includegraphics[width=\linewidth]{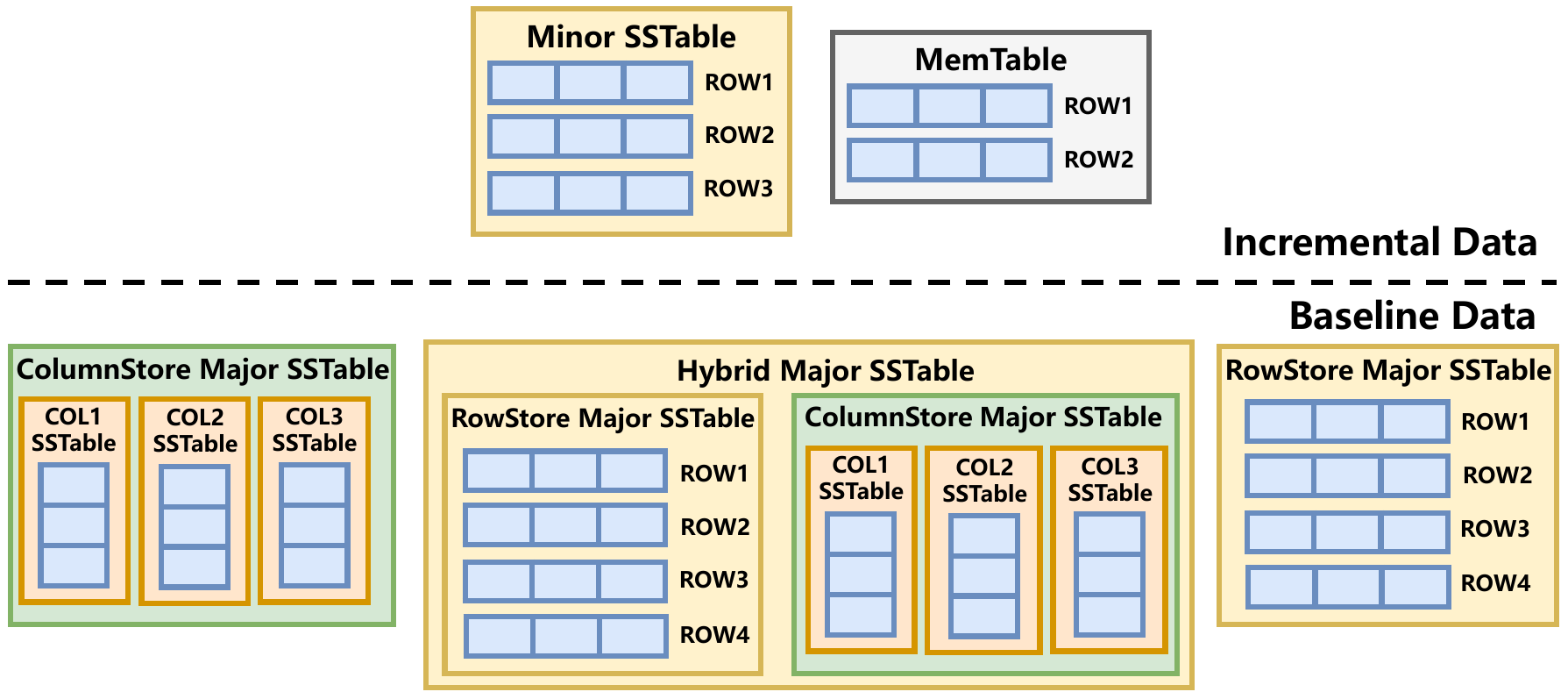}
    \vspace{-12pt}
    \caption{Hybrid Storage Architecture}
    \label{fig:HybridStorage}
    \vspace{-10pt}
\end{figure}

\noindent$\bullet\ $ In column store mode, each column data is stored as an independent SSTable, and all column SSTables are combined into a virtual SSTable as the user's column store baseline data, as shown in Figure~\ref{fig:HybridStorage}.

\noindent$\bullet\ $ Based on the user's table creation settings, baseline data can be stored in three modes: row store, column store, or redundant row store and column store.

\subsection{TP/AP Integration}

\begin{figure}[htp]
    \centering
    \includegraphics[width=0.95\linewidth]{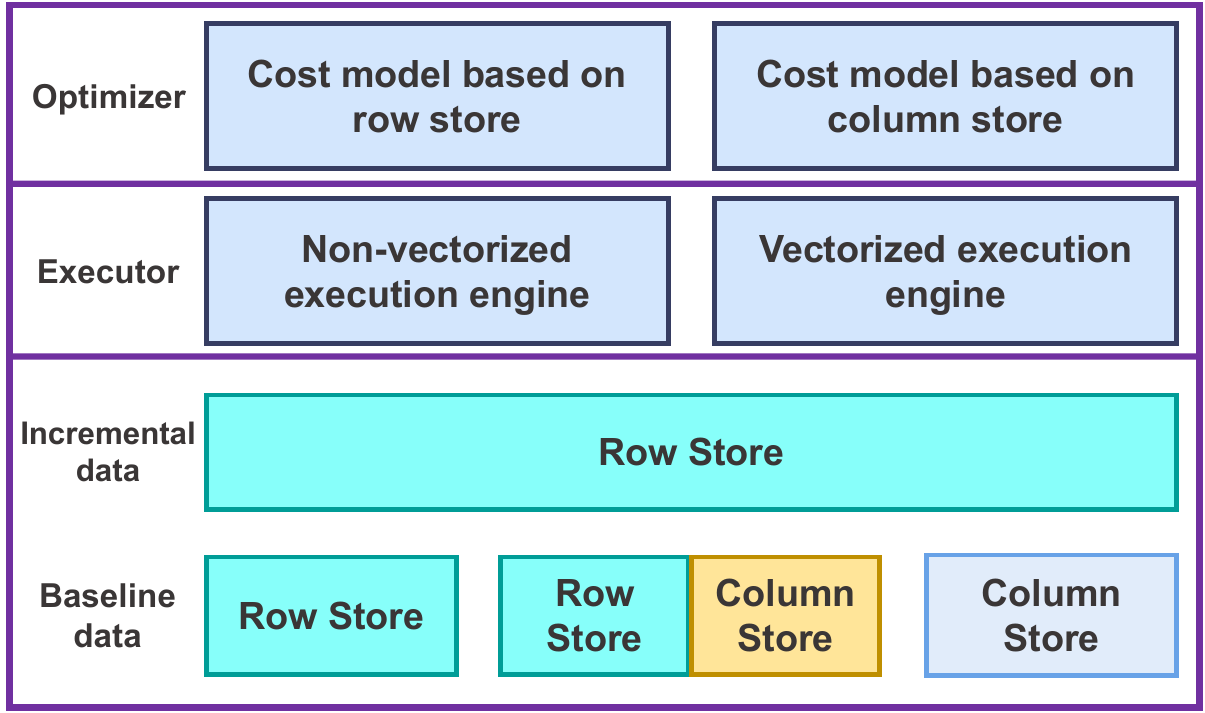}
    \vspace{-6pt}
    \caption{Row store and column store integration}
    \vspace{-6pt}
    \label{fig:RowColumnStoreIntegration}
\end{figure}

As shown in Figure~\ref{fig:RowColumnStoreIntegration}, we have not only implemented column store in the storage engine but also adapted and optimized the optimizer, executor, and other storage-related modules. This enables existing OceanBase customers with OLAP requirements to transition from row-store to column-store with no impact on their running operations. Users from other OLAP databases can also migrate their data via the migration tool (i.e., OceanBase Migration Service (OMS)~\cite{OMS}) provided by OceanBase and gain access to OceanBase's unified architecture. Users can leverage the performance benefits of column store as effortlessly as they would with row store. This integration also empowers OceanBase to achieve true TP/AP synergy, supporting diverse business workloads through a unified engine and codebase. For more details about OceanBase's TP capabilities, please refer to~\cite{DBLP:journals/pvldb/YangYHZYYCZSXYL22}.


\noindent$\bullet\ $\textbf{TP/AP integrated SQL engine.} We have developed an innovative cost model specifically tailored for columnar storage architectures, incorporating column-oriented statistical metrics into the query optimizer. The enhanced optimizer dynamically generates execution plans through cost-based analysis that accounts for the underlying data storage format. Furthermore, we have engineered a novel vectorized processing engine with architectural upgrades to core relational operators. This adaptive system intelligently modulates vectorization granularity and batch processing dimensions through continuous cost evaluation, achieving optimal resource utilization across diverse query patterns.

\noindent$\bullet\ $\textbf{TP/AP integrated storage engine.} User data is specified according to the table mode, and can be flexibly set to column store, row store or row and column redundancy mode according to the business load type. User query/backup and recovery operations are completely transparent. Column store tables fully support all online and offline DDL operations, fully support all data types and secondary index creation, ensuring that the user's usage method is no different from row store.

\noindent$\bullet\ $\textbf{TP/AP integrated transaction engine.} All incremental data is stored in rows. Transaction modifications, log content, and multi-version control all share the same logic as row store.

\subsection{Adaptive Compaction}

After the introduction of the new column store mode, the data merging behavior has changed significantly from the original row store data. Since all incremental data is row store, it needs to be merged with the baseline data, which is column stored, and split into independent SSTables for each column. The merging time and resource usage will increase significantly compared to row store.

In order to accelerate the merging speed of column store tables, the compaction process has been greatly optimized. In addition to acceleration using horizontal splitting, parallel processing, and merging, as with row store tables, vertical splitting acceleration has also been implemented for column store tables. Moreover, column store tables will combine the merging of multiple columns into a single merging task. The number of columns in one task can be autonomously determined based on the available system resources, allowing for a more balanced approach between overall merging speed and memory usage.

\subsection{Column Encoding}\label{sec: col-encoding}
 
OceanBase stores data through two levels of compression. The first level is OceanBase's self-developed row-column hybrid encoding compression, and the second level is general compression. The self-developed row-column hybrid encodings are built-in database algorithms. They can support direct query without decompression and can use encoding information to accelerate query filtering, especially for AP queries. For example, in the delta encoding, when a stored column has fixed-length numerical values, only the minimum value and the differences between each row to the minimum value are stored. It also supports string encodings like prefix encoding for columns with common prefixes. Moreover, when one column is the prefix of the other column, it stores a complete column and the suffix of another, reducing redundancy, especially for columns with repetitive timestamps. More details are available in the documentation of OceanBase\footnote{\scriptsize{https://en.oceanbase.com/docs/common-oceanbase-database-10000000001973723}}.

Since the original row-column hybrid encoding algorithms tend to be implemented for row-organized storage, OceanBase Mercury specifically designs new column encoding algorithms for columnar storage tables. Compared with the original algorithms, the new approach supports fully vectorized query execution, enables SIMD optimization across multiple instruction sets, and significantly improves compression ratios for numeric data, delivering holistic improvements in both query performance and compression ratio.

\subsection{Data Skipping Index}

Data skipping index, sometimes known as Small Materialized Aggregates~\cite{DBLP:conf/vldb/Moerkotte98} or Zone Map~\cite{DBLP:conf/dawak/Graefe09}, is a popular method for pruning irrelevant data on query processing with trivial overhead on storage and computing. Unlike traditional secondary indexes that accelerate range scans or point selects by storing index key columns in specialized data structures (e.g., B+ tree), it is orders of magnitude smaller than the raw data. 
It achieves robust performance without requiring a finely optimized table structure or precisely defined query ranges on index keys, even when optimizer-generated filters from user queries exhibit low selectivity (i.e., filter rates). 
Additionally, data skipping indexes offer greater flexibility in deployment and significantly reduced maintenance complexity compared to conventional indexing methods.


We adopted this idea to compute and store pre-aggregated data, such as the minimum value, the maximum value, sum values, and \texttt{null} counts as a sketch over data on each column. These sketches are persisted in the block-based, tree-structured block index on each column SSTable as our data skipping index. The sketches in the index tree will aggregate recursively so that data skipping index in an index tree node can represent the sketch of all children nodes it is pointing to, making data skipping index on different layers of the index tree perform as pre-aggregation of data on different granularities. This multi-granularity pre-aggregation can provide data sketch from data block granularity to SSTable granularity, and fit well with flexible range query on column and random data modification that would invalidate pre-aggregated data covering the affected data in DML. Along with the data skipping index, predicates pushed down to the storage engine can prune irrelevant data effectively without accessing all data blocks, reducing unnecessary data access overhead and I/O consumption. Sketches in a skipping index can also be used for efficient aggregation computing and data statistics collection.

In comparison to existing industrial implementations, our architecture extends the applicability of data skipping index to a broader range of scenarios. Unlike databases that separate data skipping index from data as a metadata service or independent index structure, and calculate data blocks involved in the query before execution (e.g., Snowflake), OceanBase integrates the data skipping index directly into the SSTable structure. This self-contained design simplifies data operations such as compaction, backup, migration, and DML, as all modifications are encapsulated within the SSTable. Furthermore, performing block evaluation during execution enables dynamic pruning for complex predicates involving runtime parameters or complex expressions. While conventional zonemap or data skipping index implementations typically rely on fixed-size granules, forcing a trade-off between pruning precision and overhead on storage and computing, data skipping index of OceanBase is inherently granule-adaptive. By aligning pre-aggregated sketches with the hierarchical nodes of the index block tree, we achieve multi-resolution data skipping. The finest granularity corresponds to the atomic read I/O unit (the data block), while coarser summaries are naturally formed as the tree ascends. This hierarchical structure enables efficient evaluation of complex combined boolean filters and has trivial overhead on maintenance of index. Beyond filter processing, data skipping index is also leveraged for aggregation processing and optimizer statistics collection, such as range, NDV, and sortedness estimation in OceanBase.

\subsection{Query Pushdown}

In OceanBase Mercury, the storage engine has fully supported vectorization and more pushdown support. In the column store engine, the pushdown function has been further enhanced and expanded, including:

\noindent$\bullet\ $\textbf{Pushdown of all query filters.} Filters for all queries can be pushed down to the storage layer. Additionally, depending on the type of filter, skip indexes and encoding information (refer to Section~\ref{sec: col-encoding}) can be leveraged to further accelerate query performance.

\noindent$\bullet\ $\textbf{Pushdown of common aggregate functions.} In non-group-by scenarios, the aggregate functions such as \emph{count}, \emph{max}, \emph{min}, \emph{sum}, and \emph{avg} can be pushed down to the storage engine for computation.

\noindent$\bullet\ $\textbf{Pushdown of group-by operations.} For columns with less NDV (Number of Distinct Values), the storage engine encodes the data by building an internal dictionary and storing references for each row. The group-by pushdown uses this dictionary information for significant acceleration.

\section{Materialized View}
\label{sec:MaterializedView}

A materialized view is a special view that stores the results of the view definition query. It is a method of speeding up AP queries by exchanging space for time. It retains the query results of some time-consuming queries to avoid repeated calculations of the queries.

\subsection{Overview}

Ordinary views only store the definition of the view, not the actual data. As a result, each time the view is queried, we need to execute the underlying query statement of the view again.
Materialized views include a dedicated container table that stores the results of the query defined in the view. When users query a materialized view, the database retrieves the data directly from this container table, eliminating the need for repeated calculations. 

Base tables include both baseline (columnar) and incremental (row-based) data. Materialized views are computed from base tables, reflecting both data types. Compaction converts incremental data to baseline but does not directly affect materialized views: full refresh queries the base table (both types), while incremental refresh uses mlog tracking all base table changes. Materialized views are separate container tables storing pre-computed results.
In addition, since the data of the materialized view is pre-calculated based on the query results of its definition, the content of the materialized view may become outdated when the base table data changes. To ensure the materialized view data remains up-to-date, OceanBase provides full refresh and incremental refresh mechanisms to refresh the results stored in the container table to reflect the latest status of the base table, which are discussed in Section~\ref{sec: full-refresh} and \ref{sec: incre-refresh}.


\subsection{Full refresh}\label{sec: full-refresh}

When a full refresh is performed, the database re-executes the query statement associated with the materialized view, recalculating and replacing the existing view result data. 

\begin{figure}[htp]
    \centering
    \includegraphics[width=0.75\linewidth]{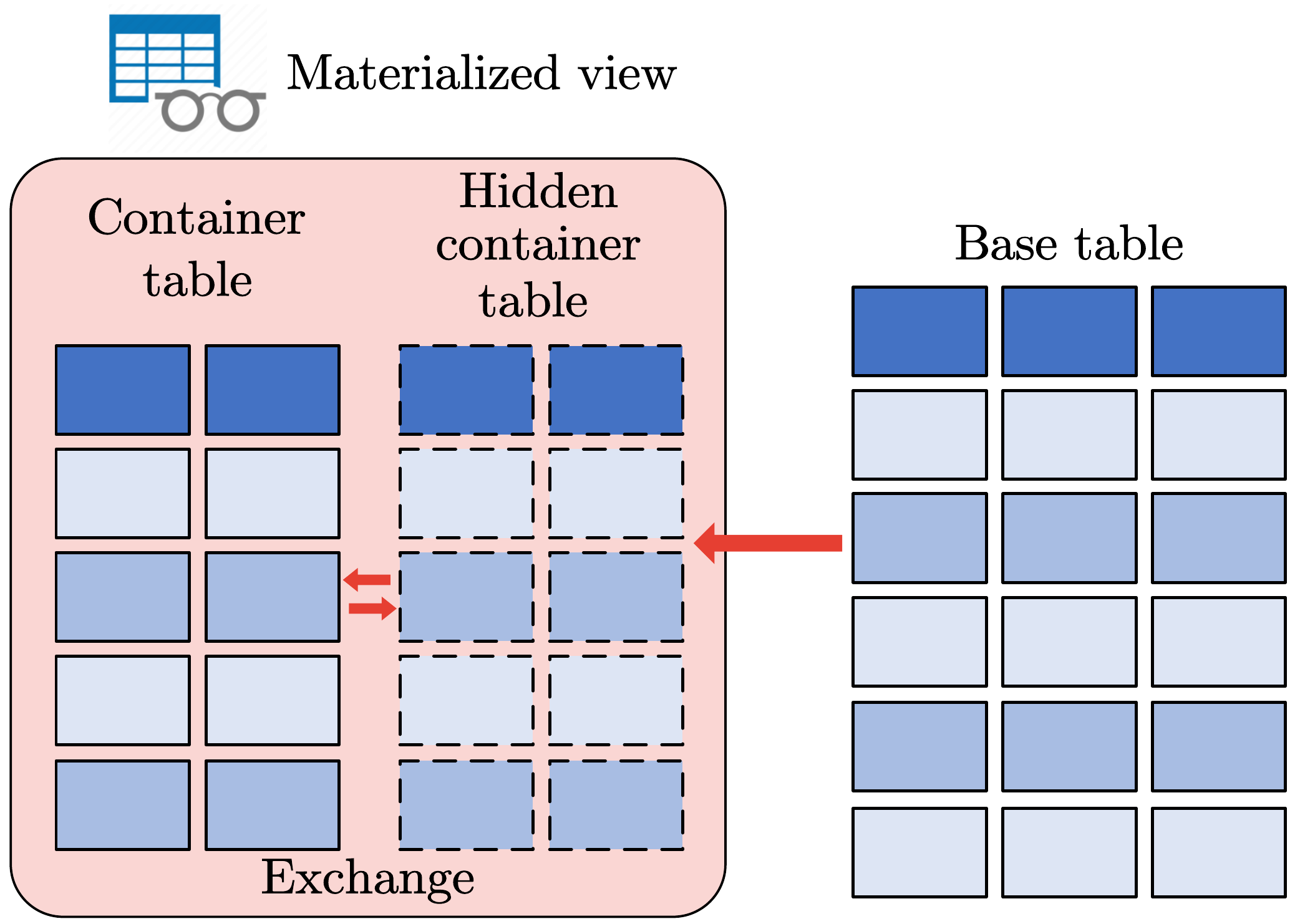}
    \caption{Full refresh}
    \vspace{-12pt}
    \label{fig:FullRefresh}
\end{figure}

OceanBase uses the off-site refresh method to perform a full refresh (as shown in Figure~\ref{fig:FullRefresh}), i.e., create a hidden table, execute a refresh statement on the hidden table, and then switch the original table and the hidden table. Therefore, the full refresh operation requires additional space and will fully rebuild the index (if any). 
A full refresh can be time-consuming and is particularly suitable for scenarios where the base table has undergone significant changes since the last refresh or when the base table contains a relatively small amount of data. Moreover, to accelerate the process, it utilizes the (full) direct load method to write the complete dataset directly into the hidden table’s data files. This method bypasses the SQL interfaces and the transaction processing layers, directly allocates space and writes the data to be inserted as SSTable structures, which is the internal representation of data in OceanBase's storage, thereby significantly improving the efficiency of data import. The full direct load method is ideal for data migration of more than 10 GB levels and for environments with limited CPU and memory resources. Its streamlined execution path minimizes CPU consumption, making it an efficient solution for such scenarios.


\subsection{Incremental refresh}\label{sec: incre-refresh}


If a materialized view is configured with an incremental refresh policy, the database periodically or immediately calculates the changes to the materialized view based on updates to the base table and applies these changes to the materialized view.
The cost of incremental refreshes is much lower than that of full refreshes and is suitable for scenarios of frequent updates of the base table. OceanBase utilizes the materialized view log (referred to as mlog later) to record the updated data. Therefore, if we want to create a materialized view with the incremental refresh policy, we need to create a materialized view log based on the base table first. The materialized view log is internally implemented as an ordinary table, which will store the old and new values of all updated rows.


\begin{figure}[htp]
    \centering
    \includegraphics[width=0.8\linewidth]{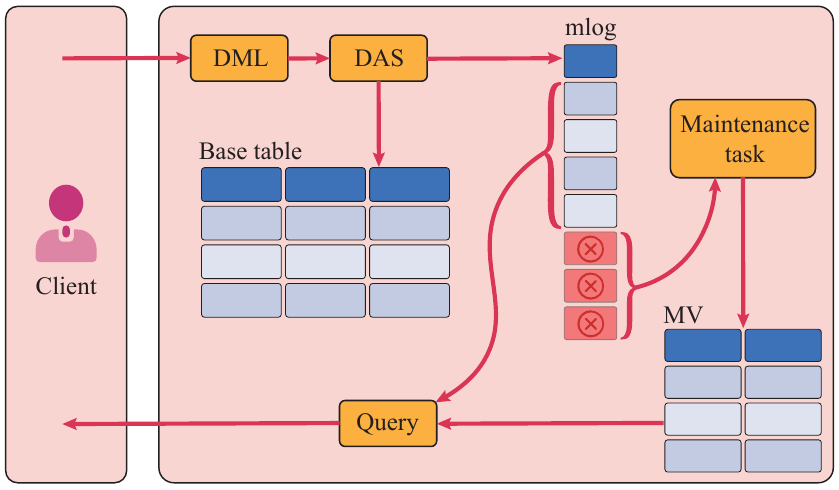}
    \caption{The overall framework of incremental update}
    \label{fig:MV}
\end{figure}

Figure~\ref{fig:MV} is the overall framework of incremental updates. For writing, all DML statements will update the base table and materialized view log simultaneously through the Data Access Service (DAS), and the background maintenance task will regularly flush the incremental updates to the materialized view. 
Besides, the maintenance task will regularly delete the records that have been updated in the materialized view. 
For querying, the \emph{Query} module will query the data in the un-updated part of MV and mlog at the same time, merge them, and return them to the user, ensuring that the queried data is the latest.

\begin{figure}[htp]
    \centering
    \includegraphics[width=0.85\linewidth]{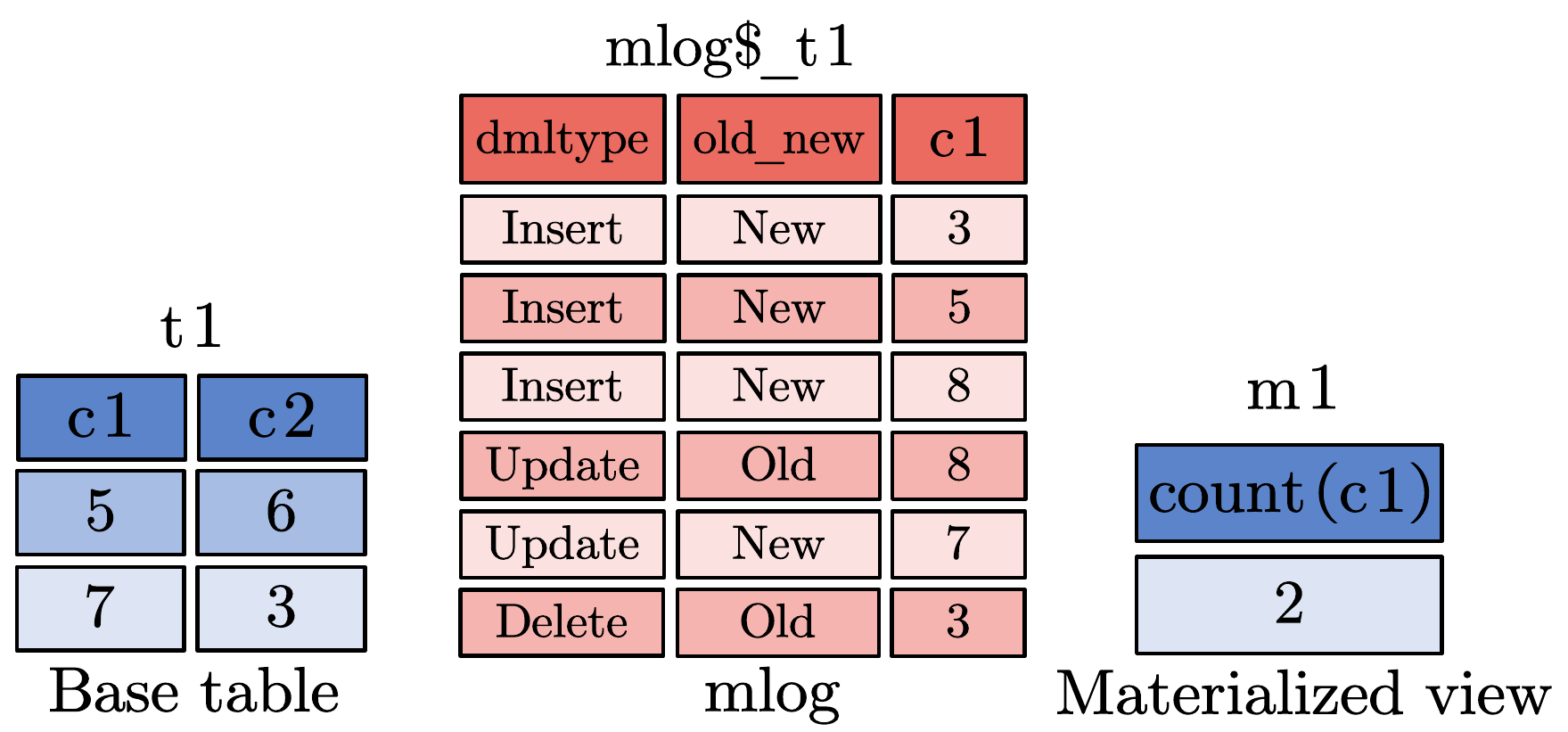}
    \caption{An example of incremental refresh}
    \label{fig:Example}
    \vspace{-6pt}
\end{figure}

We proceed to use an example to describe incremental refresh, as shown in Figure~\ref{fig:Example}. The definitions of the original tables \emph{t1} and materialized view \emph{m1} are as follows:

\begin{tcolorbox}[colback=white, colframe=black, boxrule=0.5pt, boxsep=1pt, arc=0pt, left=1pt, right=1pt, top=1pt, bottom=1pt, arc=1mm]\noindent\texttt{\small create table t1 (c1 int primary key,c2 int);\\
create\! materialized \!view m1 as\! select\! count(c1)\! from\! t1;}\end{tcolorbox}
The insert/delete/update sequence reads as follows:

\begin{tcolorbox}[colback=white, colframe=black, boxrule=0.5pt, boxsep=1pt, arc=0pt, left=1pt, right=1pt, top=1pt, bottom=1pt, arc=1mm]\noindent\texttt{\small insert into t1 values(3,4);\\
insert into t1 values(5,6);\\
insert into t1 values(8,3);\\
update t1 set c1 = 7 where c1 = 8;\\
delete from t1 where c1 = 3;}
\end{tcolorbox}


The corresponding \emph{mlog} is \emph{mlog\$\_t1}, where \emph{dmltype} specifies the type of the DML operation, and \emph{old\_new} indicates whether the row captures the data before or after the update. 
The updated value of the \emph{m1} is obtained through:
\begin{tcolorbox}[colback=white, colframe=black, boxrule=0.5pt, boxsep=1pt, arc=0pt, left=1pt, right=1pt, top=1pt, bottom=1pt, arc=1mm]\noindent\texttt{\fontsize{8pt}{10pt}\selectfont (select count(c1) from m1 where old\_new = \textquotesingle New\textquotesingle) -\\ (select count(c1) from m1 where old\_new = \textquotesingle Old\textquotesingle);}
\end{tcolorbox}


OceanBase uses the in-place refresh method for incremental refresh, i.e., executing refresh statements directly on materialized views. If there is too much incremental data, incremental refresh may be slower than full refresh. To accelerate the process, it also supports the (incremental) direct load for incremental refresh. This method directly structures the incremental data into the storage-level format (i.e., SSTable, consisting of its metadata information and a series of data) and writes the updated data directly into the storage layer. The incremental bypass import significantly enhances refresh efficiency.

\subsection{Refresh Algorithm Complexity Analysis}\label{sec:refresh-complexity}

The incremental refresh algorithm consists of two phases: (1) \emph{Refresh Type Checking} to determine the refresh strategy, and (2) \emph{Refresh DML Generation} to create SQL update statements. We analyze the complexity using the following notation: \textbf{T} (tables), \textbf{C} (columns), \textbf{A} (aggregates), \textbf{J} (JOINs), \textbf{S} (SELECT items), \textbf{D} (delta rows), \textbf{M} (MV rows), \textbf{N} (UNION ALL branches), and \textbf{P} (partitions).

Table~\ref{tab:refresh-complexity} summarizes the complexity for different view types. Simple MAV has the lowest complexity $O(S \times A)$, while Outer Join views exhibit quadratic complexity $O(T^2 \times (S + J + A))$ due to table pair processing. Major Refresh benefits from partition-level processing with $O(P \times (S + J))$ complexity. The actual execution complexity is $O(D \times \log(M))$ for refresh operations with proper indexes, or $O(D \times \log(M) / P)$ with parallel degree $P$.

\begin{table}[htp]
\centering
\caption{Refresh Algorithm Complexity Analysis}
\label{tab:refresh-complexity}
\footnotesize
\begin{tabular}{lcc}
\toprule
\textbf{View Type} & \textbf{Time} & \textbf{Space} \\
\midrule
Simple MAV & $O(S \times A)$ & $O(S + A)$ \\
Simple MJV & $O(T \times (S + J))$ & $O(T \times S)$ \\
Join MAV & $O(T \times (S + A + J))$ & $O(T \times (S + A))$ \\
Outer Join MJV/MAV & $O(T^2 \times (S + J + A))$ & $O(T^2 \times S)$ \\
UNION ALL MV & $O(N \times T \times (S + C))$ & $O(N \times T \times S)$ \\
Major Refresh MJV & $O(P \times (S + J))$ & $O(P \times S)$ \\
Execution (indexed) & $O(D \times \log(M))$ & $O(D)$ \\
\bottomrule
\end{tabular}
\end{table}

\section{Vectorized Execution Engine}
\label{sec:VectorizedExecutionEngine}

Vectorized execution is an efficient technology for batch processing of data. In analytical queries, vectorized execution can greatly improve execution performance. Starting from OceanBase 4.0~\cite{DBLP:journals/pvldb/YangXGYWZKLWX23}, the vectorized execution engine is enabled by default. Its execution performance has been greatly improved by optimizing the data format, operator implementation, and storage vectorization.

\subsection{Data Format Optimization}

\begin{figure}[htp]
    \centering
    \includegraphics[width=\linewidth]{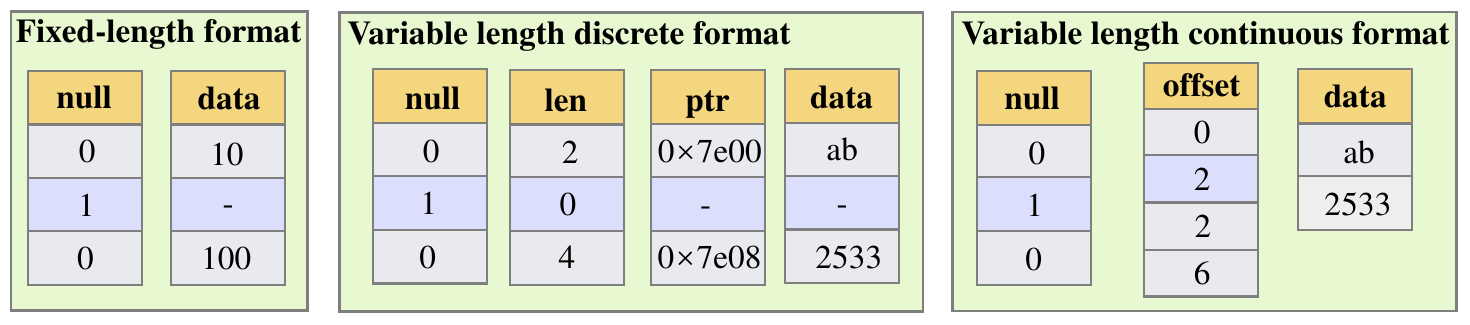}
    \vspace{-6pt}
    \caption{Data Format of Vectorized Engine in OceanBase}
    \vspace{-6pt}
    \label{fig:VectorizedEngine2}
\end{figure}

A new column-based data format is implemented in the vectorized engine of OceanBase, as shown in Figure~\ref{fig:VectorizedEngine2}. The data description information \emph{null}, \emph{len}, and \emph{ptr} information are stored separately and continuously by column to avoid redundant storage of data information. For different data types and usage scenarios, three data formats are implemented: fixed-length data format, variable-length discrete format, and variable-length continuous format:

\noindent$\bullet\ $\textbf{Fixed-length data format} only requires a \emph{null} bitmap and continuous data information, and only one length value needs to be stored. It is not necessary to store the same length value for each data in a batch of data redundantly, and pointer information for indirect access is no longer required. 
Data information is not stored redundantly, which saves more space; it can be directly accessed, and the access data locality is better; data can be stored continuously, which is more friendly to SIMD instructions. In addition, when materializing and serializing, there is no need to perform pointer swizzling operations on data, which is more efficient.

\noindent$\bullet\ $\textbf{Variable-length discrete format} means that in a batch of data, each data may be stored discontinuously in memory, and each data is stored continuously in columns using data address pointer and length description, length information, and pointer information; using this format, if the storage layer is encoded data, it does not need to deeply copy the data during projection, only \emph{len} and \emph{ptr} information needs to be projected. For short-circuit computing scenarios, a batch of data may only calculate a few rows, and this format description can also be used without reorganizing the data.

\noindent$\bullet\ $\textbf{Variable-length continuous format} means that the data is stored continuously in the memory. The length information and offset address of each data are described using the offset array. Compared with the discrete format, this description format needs to ensure data continuity when organizing data. It is more efficient for data access and data batch copying. However, it is not very friendly for short-circuit computing scenarios and column-encoded data projection, where the data is reorganized and deeply copied. Currently, this format is mainly used in column-based materialization scenarios.

Upon reading micro-block data from the disk, the storage layer proceeds to decompress and decode the data before projecting it to the memory format corresponding to the SQL layer. The proposed fixed-length format offers enhanced performance compared to the previous format. Notably, the new fixed-length format streamlines the process by eliminating the need to set \textit{len} and \textit{ptr} when projecting the memory format, and allows for efficient batch \textit{memcpy} operations due to the continuous storage of fixed-length data. This results in improved efficiency compared to the previous format, where filling a batch of fixed-length data required setting length, \textit{null}, and \textit{ptr} for each Datum data, as well as filling the integer data value into the data buffer pointed to by \textit{ptr}.


\subsection{Operator and Expression Optimization}

The vectorized engine has comprehensively optimized the implementation of operators and expressions. The main optimization idea is to use batch data attribute information, algorithm and data structure optimizations, and specialized implementation based on the new format, thereby reducing CPU data cache misses, CPU branch prediction errors, and CPU instruction overhead, and improving overall execution performance. The vectorized engine has redesigned and implemented operators and expressions such as Sort, Hash Join, Hash Group By, Data Shuffle, and Aggregate Calculation in the new format, and the overall computing performance has been comprehensively improved.

\subsubsection{Using batch data attribute information}

The vectorized engine maintains the characteristic information of batch data during execution, including whether \emph{null} exists and all rows in the batch need not be filtered. The information greatly speeds up expression calculation. For example, if \emph{null} does not exist, there is no need to consider special treatment of \emph{null} during expression calculation. If no rows are filtered, there is no need to determine whether each row has been filtered during calculation. In addition, if the data is continuous and not filtered, it is more friendly to SIMD calculation.

\subsubsection{Algorithm and data structure optimizations}

In terms of algorithm and data structure optimizations, a more compact materialization structure is implemented, supporting data materialization by row/column, saving space and making access more efficient; the \emph{Sort} operator realizes the separate materialization of sort key and non-sort key, combined with the sort key sequence-preserving encoding (encoding multiple columns of data into one column, which can be directly compared using \emph{memcmp}), the data cache miss that the \emph{Sort} operation accesses is lower during the comparison process, the comparison calculation is faster, and the overall sorting efficiency is higher. \emph{HashGroupBy} optimizes the Hash table structure, and the data storage in HashBucket is more compact. The low-cardinality Group Key is optimized using \emph{array}, and the grouping and aggregation results are stored in memory continuously, etc.

\subsubsection{Specialized implementation optimization}

It mainly uses templates to generate more efficient implementations for different scenarios. For example, the Hash Join specialization implements the encoding of multiple fixed-length join keys into one fixed-length column and puts the join key data into the bucket. It also optimizes data expectations and reduces data cache misses when accessing multiple columns of data. It supports specialized implementation of aggregate calculations, and implements different aggregate calculations separately, thereby reducing the number of aggregate function instructions and branch judgments for each calculation, and greatly improving execution efficiency.

\subsubsection{Storage vectorization optimization}

The storage layer fully supports the new vectorization format, and SIMD is used more for projection, predicate pushdown, aggregation pushdown, and \emph{group by} pushdown. When projecting fixed-length and variable-length data, the template is customized according to the column type, column length, and whether it contains \emph{null} information, the projection is shallowly copied in batches. When calculating the pushdown predicate, for simple predicate calculations, it is directly performed on the column encoding; for complex predicates, it is projected into a new vectorization format and calculated in batches on the expression. Aggregation pushdown makes full use of the pre-aggregation information of the middle layer, such as count, sum, max, min, etc. \emph{Group by} pushdown makes full use of the encoded data information, and the acceleration effect is very noticeable for dictionary-type encoding.


\section{Experiments}
\label{sec:evaluation}
In this section, we conduct extensive experiments to evaluate OceanBase Mercury (an alias for OceanBase 4.3.3) under different configurations and compare it with various open source OLAP systems. 

\subsection{Experimental Setup}

\noindent\textbf{Hardware settings.} We deployed OceanBase Mercury both on a single server and on a distributed cluster comprising three distinct servers. 
Unless otherwise stated, the experiments were evaluated on servers with a 128-core Intel(R) Xeon(R) Platinum 8369B CPU@2.90GHz, 1TB DRAM, and 44TB HDD.

\noindent\textbf{Baselines.} We conducted a comprehensive comparison of OceanBase 4.3.3 and other distributed database systems, namely StarRocks 3.3.3 and ClickHouse 24.10.1. To ensure a fair evaluation, each of these systems was deployed in the same region, with an identical scale of clusters.

\noindent\textbf{Benchmarks.} In the experiments, we employed three kinds of benchmarks: TPC-H~\cite{tpch}, TPC-DS~\cite{tpcds}, and ClickBench~\cite{ClickBench}.

\noindent\textbf{Metrics.} We evaluated the performance of the databases by measuring the query latency of benchmark queries. Each SQL statement is executed under both hot run and cold run conditions to assess its query latency. For a cold run, we restart the database system and clear the memory before executing the queries and recording their execution times. In contrast, for a hot run, we execute each SQL statement three times and record the shortest execution time.

\subsection{Performance evaluation of column encoding}

\begin{figure}[htp]
    \centering
    \includegraphics[width=1.0\linewidth]{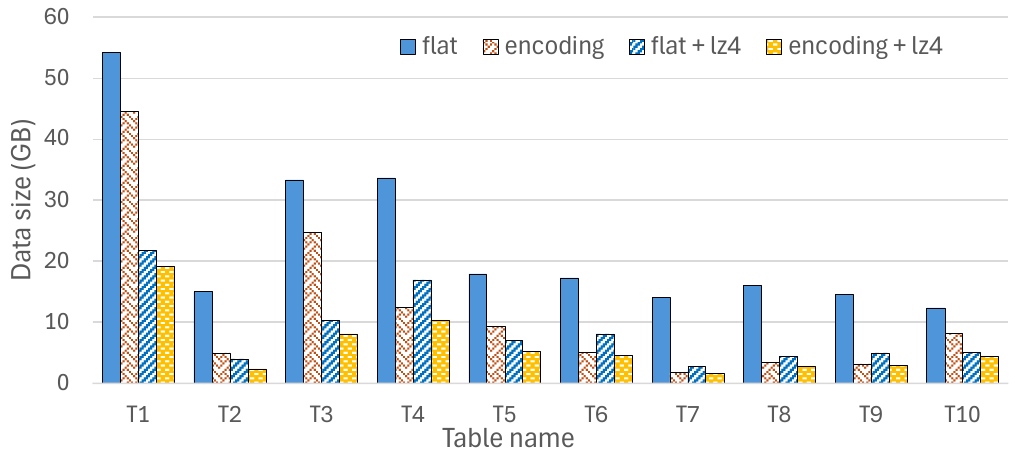}
    \caption{Data size after encoding}
    \label{fig:encoding}
    \vspace{-6pt}
\end{figure}

OceanBase Mercury implements advanced column encoding techniques, including multi-prefix, inter-column equality, and inter-column substring encoding, to improve compression ratios and reduce storage overhead. As shown in Figure~\ref{fig:encoding}, ten tables (i.e., $T1$, $\cdots$, $T10$) from a practical business scenario are used in our experiment. Adding multi-prefix, inter-column equality, and inter-column substring encoding improved compression ratios for about half of the tables. $T1$'s space savings rose from 15\% to 17\% (LZ4: 10\% to 12\%), mainly due to prefix and inter-column substring encoding. $T2$ and $T3$ increased from 59\% to 67\% and 16\% to 25\%, respectively, primarily via prefix encoding. $T5$ saw a significant gain—from 30\% to 48\% (LZ4: 20\% to 26\%) using all three techniques. $T7$ achieved the largest improvement, jumping from 66\% to 87\% (LZ4: 32\% to 43\%), driven by prefix encoding. $T10$ also improved modestly, from 28\% to 33\% (LZ4: 10\% to 11\%). These results highlight the effectiveness of exploiting column-level redundancies for better compression.

\subsection{Performance comparison of row-based and column-based materialized views}

In the current database environment, materialized views, as an optimization method, have been widely used to accelerate report generation and data analysis, especially in scenarios where multiple tables need to be joined to build a large wide table for in-depth analysis. At the same time, OceanBase has successfully achieved the integration of row storage and column store since version 4.3.0 by integrating a new column store engine, which not only greatly enhances the query performance in OLAP scenarios, but also takes into account the needs of OLTP services. 

However, OceanBase's materialized views used a row storage format before this version. Although it can meet basic data storage requirements, its performance improvement space is limited when facing large amounts of data and complex analytical queries. To make up for this shortcoming, especially in OLAP scenarios dominated by analytical queries such as OLAP systems and data warehouses, the introduction of column store materialized views to process large data sets and perform complex analytical tasks can improve query efficiency.
We performed this experiment in a distributed cluster consisting of three servers.

\begin{table*}[ht]
    \caption{Latency of queries over Row-based MV v.s. Col-based MV across varying row counts and base table storage engines.}
    \centering
    \vspace{-3mm}
    \begin{tabular}{p{26pt}<{\centering}|p{12pt}<{\centering}|p{30pt}<{\centering}|p{30pt}<{\centering}|p{12pt}<{\centering}|p{30pt}<{\centering}|p{30pt}<{\centering}|p{12pt}<{\centering}|p{30pt}<{\centering}|p{30pt}<{\centering}|p{12pt}<{\centering}|p{30pt}<{\centering}|p{30pt}<{\centering}}
        \bottomrule[1pt]
        \multirow{3}{*}{\!\!\!\!\!\!\!Operators\!\!\!\!\!\!} & \multicolumn{6}{c|}{\rule{0pt}{9pt}Base table with $10^8$ rows (ms)} & \multicolumn{6}{c}{\rule{0pt}{9pt}Base table with $10^9$ rows (ms)}\\
        \cline{2-13}
        & \multicolumn{3}{c|}{\rule{0pt}{9pt}Row-stored base table} & \multicolumn{3}{c|}{Col-stored base table} & \multicolumn{3}{c|}{Row-stored base table} & \multicolumn{3}{c}{Col-stored base table}\\
        \cline{2-13}
        & \multicolumn{1}{c|}{\rule{0pt}{9pt}\!\!View\!\!} & \multicolumn{1}{c|}{\rule{0pt}{9pt}\!\!Rbased MV\!\!} & \multicolumn{1}{c|}{\!\!\!Cbased MV\!\!\!} & \multicolumn{1}{c|}{\rule{0pt}{9pt}\!\!View\!\!} & \multicolumn{1}{c|}{\!\!Rbased MV\!\!} & \multicolumn{1}{c|}{\!\!\!Cbased MV\!\!\!} & \multicolumn{1}{c|}{\rule{0pt}{9pt}\!\!View\!\!} & \multicolumn{1}{c|}{\!\!Rbased MV\!\!} & \multicolumn{1}{c|}{\!\!\!Cbased MV\!\!\!} & \multicolumn{1}{c|}{\rule{0pt}{9pt}\!\!View\!\!} & \multicolumn{1}{c|}{\!\!Rbased MV\!\!} & \multicolumn{1}{c}{\!\!\!Cbased MV\!\!\!}\\
        \cline{1-13}
        \rule{0pt}{9pt}\!count(*)\!   & 32.3   & 4.7   & 3.8   & 13.9   & 6.3   & 3.7 & 31.9   & 5.1   & 5.0   & 13.9   & 6.2   & 3.7\\
        \rule{0pt}{9pt}\!\!\!\!count(c1)\!\!\!\!  & 32.6   & 2.6   & 2.6   & 13.8   & 3.3   & 2.5 & 31.6   & 2.3   & 2.0   & 13.7   & 3.0   & 2.7\\
        \rule{0pt}{9pt}\!\!\!\!count(c2)\!\!\!\!  & 32.2   & 2.6   & 2.1   & 13.8   & 3.0   & 2.5 & 31.8   & 2.0   & 2.0   & 13.9   & 3.0   & 2.3\\
        \rule{0pt}{9pt}\!\!sum(c2)\!\!    & 32.8   & 2.7   & 2.4   & 13.9   & 3.0   & 2.6  & 32.6   & 2.0   & 2.0   & 13.6   & 3.0   & 2.2\\
        \rule{0pt}{9pt}avg(c2)    & 32.3   & 2.5   & 2.2   & 13.8   & 3.0   & 2.4 & 32.7   & 2.0   & 1.8   & 13.8   & 3.0   & 2.2\\
        \rule{0pt}{9pt}\!\!max(c2)\!\!    & 32.7   & 2.5   & 2.0   & 13.9   & 3.0   & 2.5 & 32.4   & 2.0   & 1.8   & 13.8   & 3.1   & 2.9\\
        \rule{0pt}{9pt}\!\!min(c2)\!\!    & 32.2   & 2.5   & 2.1   & 13.8   & 3.2   & 2.6 & 33.2   & 3.0   & 1.7   & 13.8   & 3.1   & 3.0\\
        \cline{1-13}
        \rule{0pt}{9pt}Total      & \!227.1   & 20.1   & 17.2   & 96.8   & 24.8   & 18.8 & \!226.2   & 18.6   & 16.3   & 96.7   & 24.4   & 19.0\\
        \toprule[1pt]
    \end{tabular}
    \label{tab:MV_table}
\end{table*}


Table~\ref{tab:MV_table} presents a comprehensive performance comparison of View queries, row-based materialized views (MV), and column-based MVs across different base table storage engines (row-stored and column-stored) and data scales ($10^8$ and $10^9$ rows). The evaluation covers seven common aggregate operations: \texttt{count(*)}, \texttt{count(c1)}, \texttt{count(c2)}, \texttt{sum(c2)}, \texttt{avg(c2)}, \texttt{max(c2)}, and \texttt{min(c2)}.

The results demonstrate that materialized views provide substantial performance improvements over direct View queries. For row-stored base tables, View queries exhibit latencies around 32--33 ms, while row-based MVs reduce this to 2--5 ms (approximately 6--11$\times$ speedup), and column-based MVs achieve 1.7--3.8 ms (approximately 8--19$\times$ speedup). For column-stored base tables, View queries show latencies around 13--14 ms, while row-based MVs achieve 3--6 ms (approximately 2--4$\times$ speedup), and column-based MVs reach 1.7--3.7 ms (approximately 4--8$\times$ speedup). Column-based MVs consistently outperform row-based MVs across all scenarios, with the total query latency reduced from 227.1 ms (View) to 17.2 ms (column-based MV) for row-stored base tables with $10^8$ rows, representing a 13$\times$ improvement. The performance advantage of column-based MVs becomes more pronounced for analytical queries involving column-specific operations, particularly \texttt{count(c1)} and \texttt{count(c2)}, where column-based MVs achieve sub-millisecond latencies (1.7--2.7 ms) compared to row-based MVs (2.0--3.2 ms). Notably, the performance remains stable as data scales from $10^8$ to $10^9$ rows, indicating excellent scalability of the materialized view implementation.

Both row-based and column-based real-time materialized views query the latest incremental data by merging mlog and MV data, achieving effectively 0 freshness. Resource consumption differs based on workload characteristics: column store MVs excel in OLAP scenarios with analytical queries, while row store MVs are more efficient for OLTP workloads. For $10^8$ rows ($\approx$ 4.3 GB without encoding and compression), the data volume is around 3.2 GB, while for $10^9$ rows ($\approx$ 43 GB without encoding and compression), the data volume is around 31 GB in OceanBase with encoding and compression.

\subsection{Performance evaluation of the vectorized execution engine}

OceanBase Mercury implements the vectorized engine, which designs new data format descriptions for various columnar data formats. 
The vectorized engine eliminates the serialization overhead and read/write access overhead associated with maintaining the previous data description approach. 
Based on the new data format description, various operators and expressions have been reimplemented, enhancing its efficiency for large-scale data processing.

\begin{figure*}[htp]
    \centering
    \includegraphics[width=\linewidth]{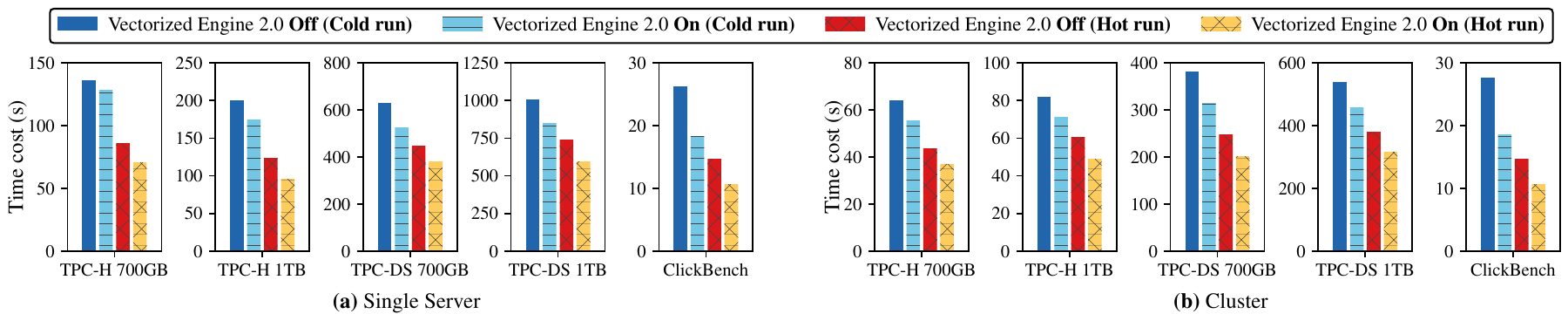}
    \caption{Total query latency of OceanBase Mercury when turning off \& turning on the vectorized engine.}
    \label{fig:vector_total}
    \vspace{-0pt}
\end{figure*}

Figure~\ref{fig:vector_total} compares the total query latency of OceanBase Mercury with the vectorized engine disabled and enabled under TPC-H (700GB, 1TB), TPC-DS (700GB, 1TB), and ClickBench benchmarks. With the vectorized engine enabled, OceanBase Mercury achieves notable improvements in both single-server and cluster settings. The vectorized engine reduces the overall query latency by 6\% to 33\% in cold run settings and 18\% to 28\% in hot run settings. The largest improvements appear in ClickBench, which features queries over a single big table. With the vectorized engine enabled, OceanBase Mercury reduces the query latency by 30\% (cold run) and 28\% (hot run) in the single server setting in Figure~\ref{fig:vector_total} (a) - ClickBench, and reduces the query latency by 33\% (cold run) and 27\% (hot run) in the cluster setting in Figure~\ref{fig:vector_total} (b) - ClickBench.

\begin{figure*}[htp]
    \centering
    \includegraphics[width=\linewidth]{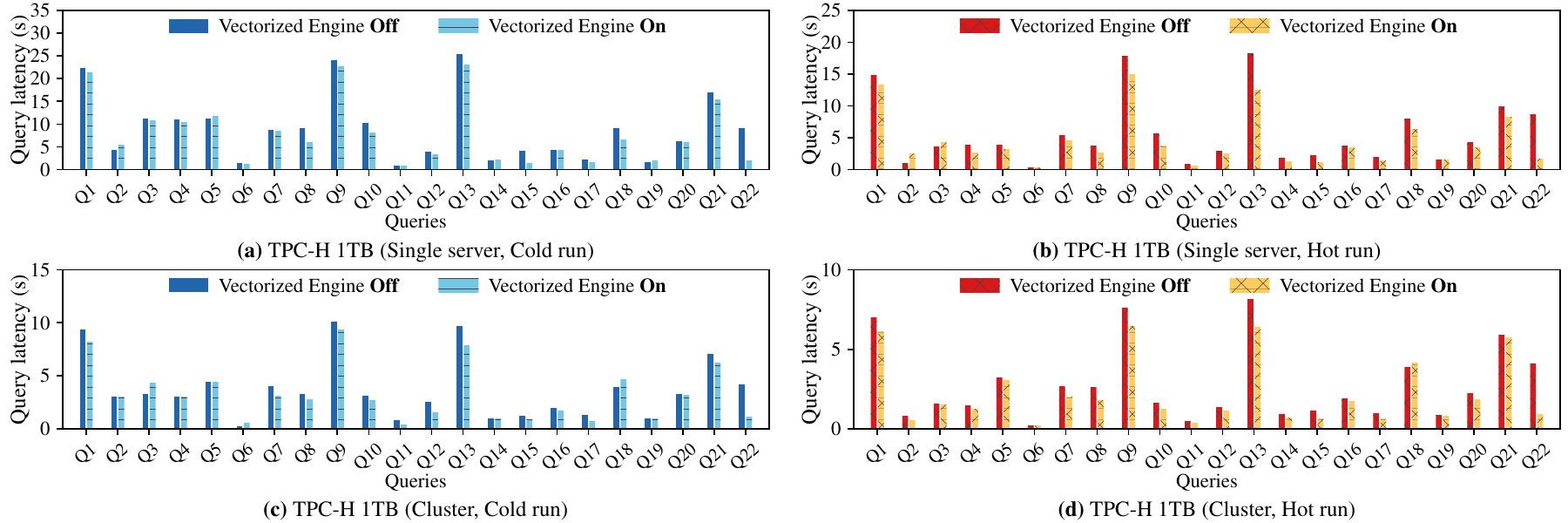}
    \caption{Under TPC-H (1TB), the query latency of OB Mercury when turning off \& turning on the vectorized engine.}
    \label{fig:vector_tpch_1t}
\end{figure*}

\begin{figure*}[htp]
    \centering
    \includegraphics[width=\linewidth]{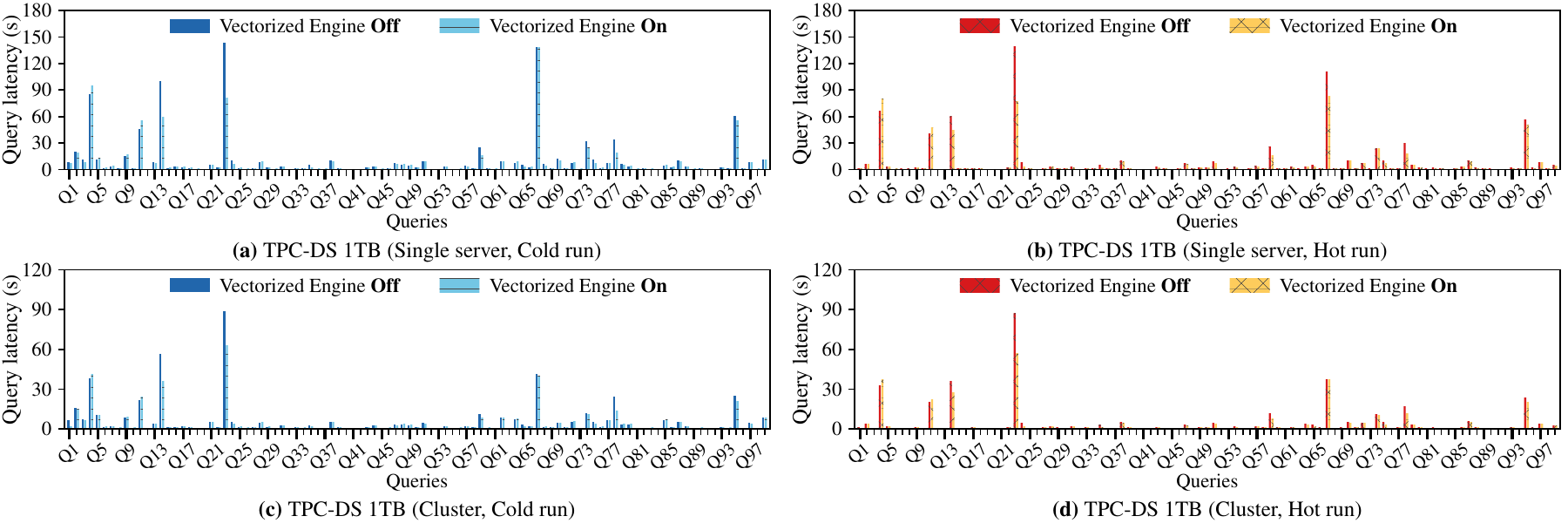}
    \caption{Under TPC-DS (1TB), the query latency of OceanBase Mercury when turning off \& turning on the vectorized engine.}
    \label{fig:vector_tpcds_1t}
\end{figure*}

\begin{figure*}[htp]
    \centering
    \includegraphics[width=\linewidth]{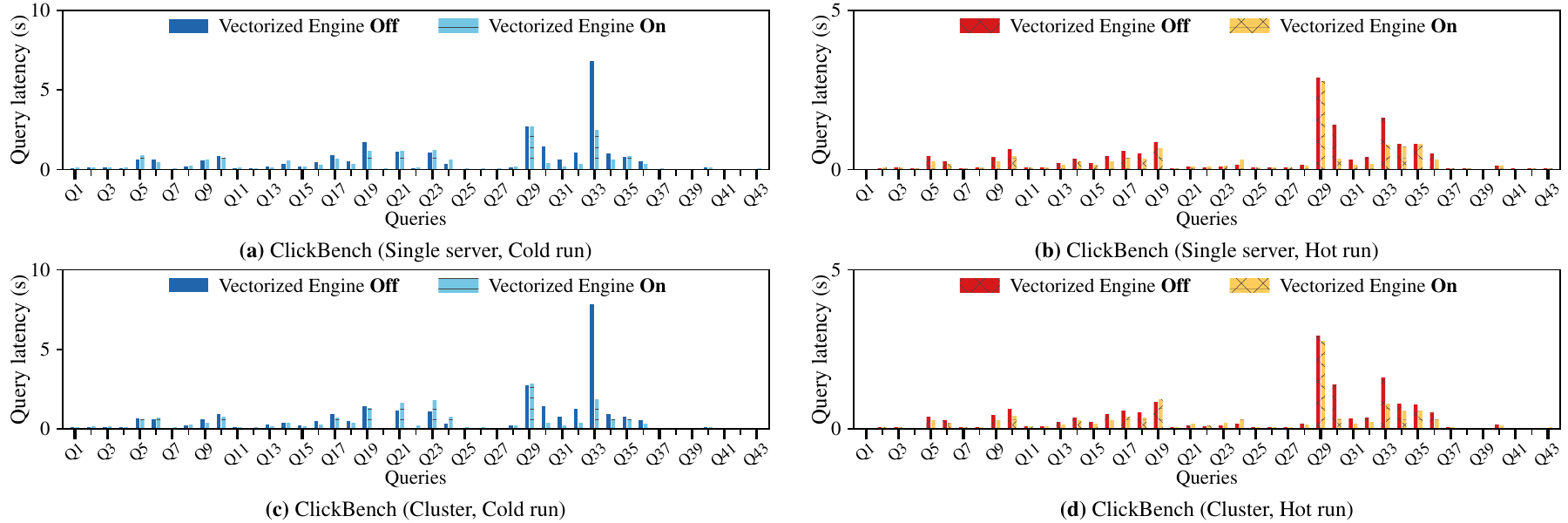}
    \vspace{-8pt}    
    \caption{Under ClickBench, the query latency of OceanBase Mercury when turning off \& turning on the vectorized engine.}
    \label{fig:vector_click}
    \vspace{-10pt}
\end{figure*}

Figure~\ref{fig:vector_tpch_1t}, Figure~\ref{fig:vector_tpcds_1t}, and Figure~\ref{fig:vector_click} present the detailed performance of each query in the TPC-H (1TB), TPC-DS (1TB) and ClickBench benchmarks, respectively. With the vectorized engine enabled, OceanBase Mercury reduces the query latency for nearly every query. The improvements are particularly notable for some of the most time-consuming queries. For example, in Figure~\ref{fig:vector_tpcds_1t} (b), the query latency for Query 23 decreases from 138.99 seconds to 77.17 seconds, achieving an improvement of approximately 44.5\%.

To further evaluate the effectiveness of the vectorized engine, we compare the performance of column store and row store configurations with vectorization 2.0 enabled. As shown in Table~\ref{tab:vectorization2_performance}, column store exhibits superior performance compared to row store across all benchmarks, with execution times of 14.54s, 19.42s, and 108.19s for ClickBench, TPCH-100G, and TPCDS-100G, respectively, representing 1.7$\times$ to 1.8$\times$ speedup over row store.

The vectorized engine 2.0 excels in computation-intensive queries. However, minor regressions occur in specific cases like ClickBench Q5 in Figure~\ref{fig:vector_click} (a) (distinct count on "UserID" column). These stem from initialization overheads, such as batch buffer allocation, which cannot be fully amortized by the computational savings in I/O-bound scenarios. Despite this, the engine delivers substantial and consistent improvements for the vast majority of analytical workloads.

\begin{table}[htp]
\centering
\caption{Performance comparison with vectorization 2.0 enabled (Hot run)}
\label{tab:vectorization2_performance}
\begin{tabular}{lcc}
\toprule
\textbf{Benchmark} & \textbf{Column Store (s)} & \textbf{Row Store (s)} \\
\midrule
ClickBench & 14.54 & 24.32 \\
TPCH-100G & 19.42 & 35.27 \\
TPCDS-100G & 108.19 & 155.40 \\
\bottomrule
\end{tabular}
\end{table}

\subsection{OceanBase (OB) v.s. StarRocks (SR) \& ClickHouse (CH)}

\begin{figure}[htp]
    \centering
    \includegraphics[width=1.0\linewidth]{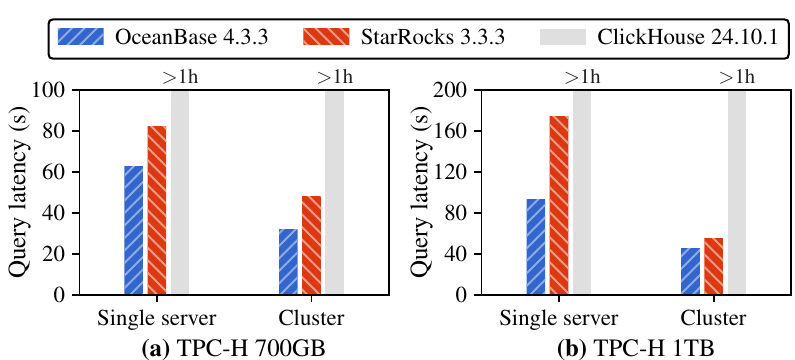}
    \caption{Total query latency of OB v.s. SR \& CH under TPC-H benchmark.}
    \label{fig:tpch_total_time}
    \vspace{-0pt}
\end{figure}

\begin{figure}[htp]
    \centering
    \includegraphics[width=1.0\linewidth]{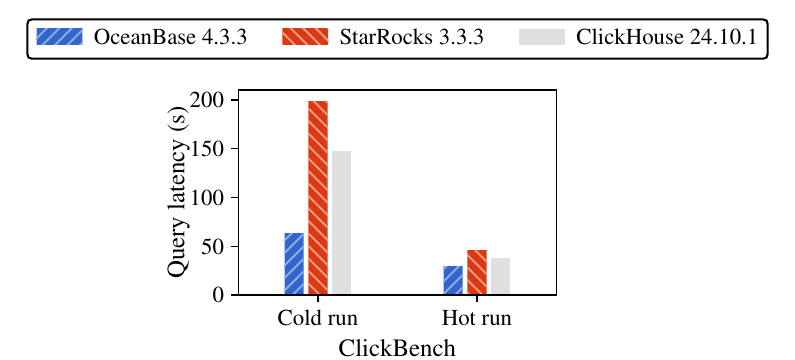}
    \caption{Total query latency of OB v.s. SR \& CH under CB benchmark.}
    \label{fig:clickbench_total_time}
    \vspace{-0pt}
\end{figure}

ClickHouse~\cite{ClickHouse, schulze2024clickhouse} and StarRocks~\cite{StarRocks} are prominent open-source column-oriented databases specifically designed for near real-time analytical processing. As public benchmarks~\cite{ClickBench_vs} indicate that ClickHouse outperforms commercial alternatives like Databricks and Snowflake, a comparison of OceanBase with these two high-performance open-source systems is sufficient.
While they are renowned for their high performance, OceanBase 4.3.3 demonstrates superior analytical processing performance compared with their latest versions. 
Figure~\ref{fig:tpch_total_time} and Figure~\ref{fig:clickbench_total_time} show the total query latency of OceanBase 4.3.3 compared with StarRocks 3.3.3 and ClickHouse 24.10.1 under TPC-H (700GB and 1TB) and ClickBench benchmarks. OceanBase 4.3.3 consumes less time than StarRocks and ClickHouse for both of the two benchmarks. 

\begin{figure*}[htp]
    \centering
    \includegraphics[width=\linewidth]{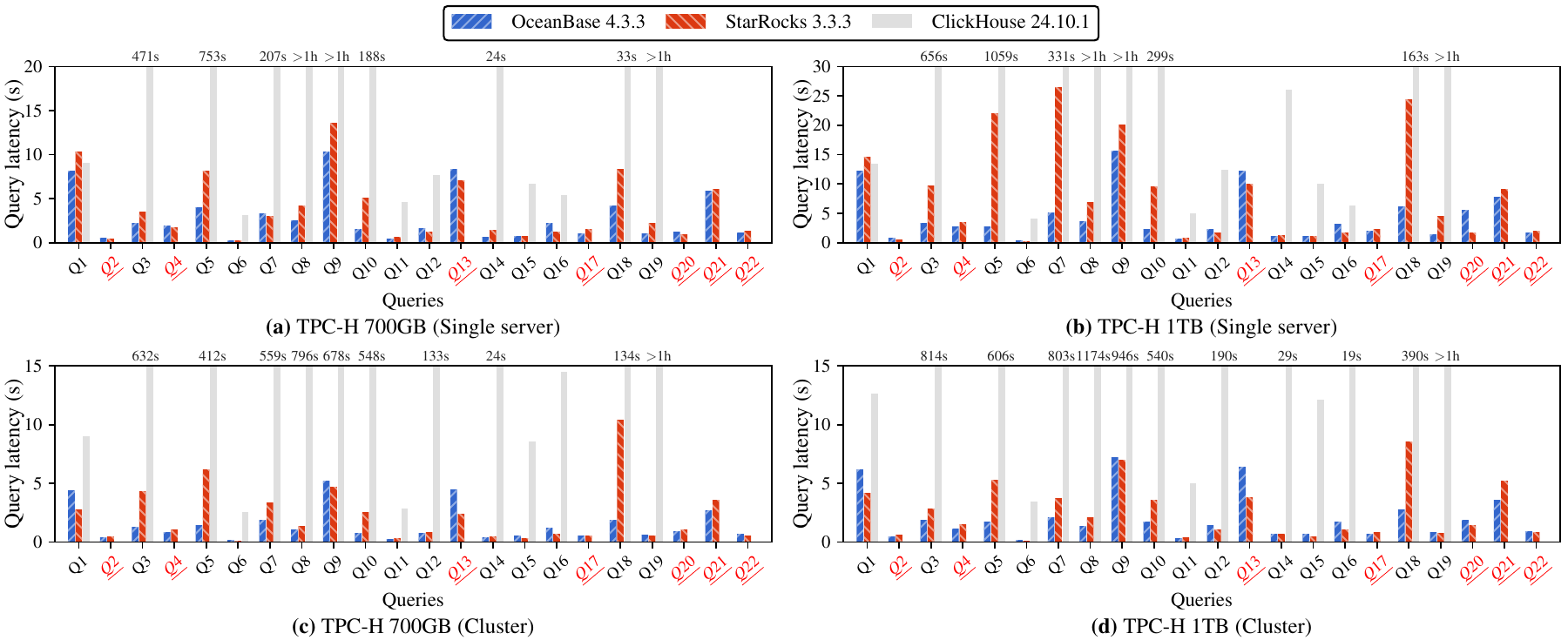}
    \caption{OB v.s. SR \& CH under TPC-H benchmark. ClickHouse does not support queries highlighted in red, so these queries are not plotted for ClickHouse.}
    \label{fig:OB_vs_SR_CK}
\end{figure*}

For the TPC-H benchmark, since ClickHouse does not support all the queries in TPC-H~\cite{schulze2024clickhouse, TPCH-ClickHoue} and takes more than one hour to process those supported queries, it is not discussed further. Compared with StarRocks, OceanBase reduces the total query latency by 23.7\% on a single server and by 33.8\% in a cluster for the 700GB TPC-H. Additionally, it achieves query latency reductions ranging from 46\% on a single server to 19\% in a cluster for the 1TB TPC-H. These improvements in query efficiency are essential for near real-time analytical processing systems. Figure~\ref{fig:OB_vs_SR_CK} displays the detailed query latency for each query in TPC-H. For those queries not supported by ClickHouse~\cite{schulze2024clickhouse}, we color their x-axis labels in red and do not plot them in the figure. OceanBase 4.3.3 achieves less query latency for most queries, especially for those complex and time-consuming queries. For instance, for Query 18 in Figure~\ref{fig:OB_vs_SR_CK} (b), OceanBase costs 6.07 seconds, while StarRocks costs 24.3 seconds and ClickHouse costs 163.0 seconds. Query 18 fetches details about suppliers with supply quantities exceeding a specified amount. The query involves a three-table join operation that simultaneously performs operations such as aggregation, grouping, sorting, and utilizing an \texttt{IN} subquery. OceanBase's superiority in the query demonstrates its capability to speed up complex queries.

\begin{figure*}[htp]
    \centering
    \includegraphics[width=\linewidth]{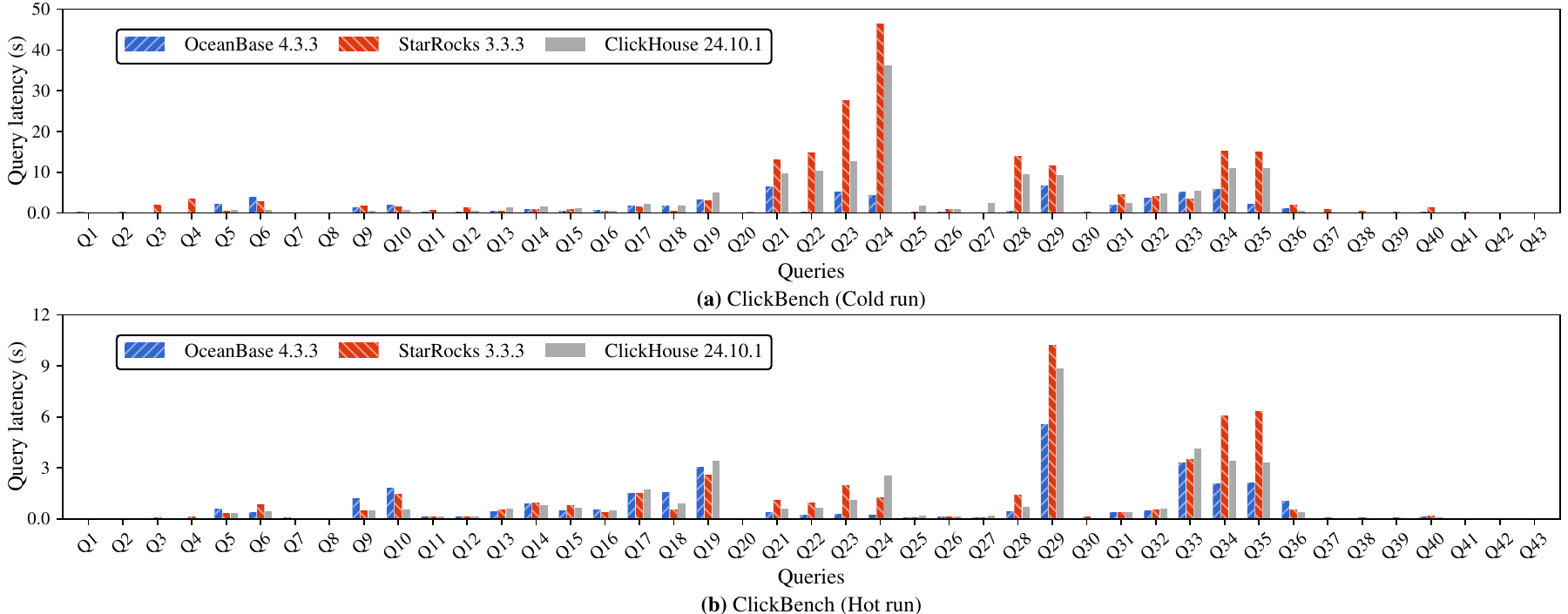}
    \vspace{-10pt}
    \caption{OceanBase v.s. StarRocks \& ClickHouse in a single server under ClickBench benchmark.}
    \label{fig:clickbench_single_cold_hot}
    \vspace{-12pt}
\end{figure*}

The performance results on the ClickBench benchmark are executed on a server with a 16-core Processor and 32GB DRAM. As shown in Figure~\ref{fig:clickbench_total_time}, OceanBase reduces the query latency by 68\% in the cold run and 35\% in the hot run compared with StarRocks, and by 57\% in the cold run and 23\% in the hot run compared with ClickHouse.
Figure~\ref{fig:clickbench_single_cold_hot} shows the detailed query latency for all queries in ClickBench, and OceanBase achieves superior performance. The most prominent improvement appears in Query 24 of Figure~\ref{fig:clickbench_single_cold_hot} (a), where OceanBase consumes 4.4 seconds while StarRocks costs 46.5 seconds and ClickHouse costs 36.2 seconds.


\subsection{OceanBase v.s. Doris on Update-Intensive Workloads}

\begin{figure}[htp]
    \centering
    \includegraphics[width=0.9\linewidth]{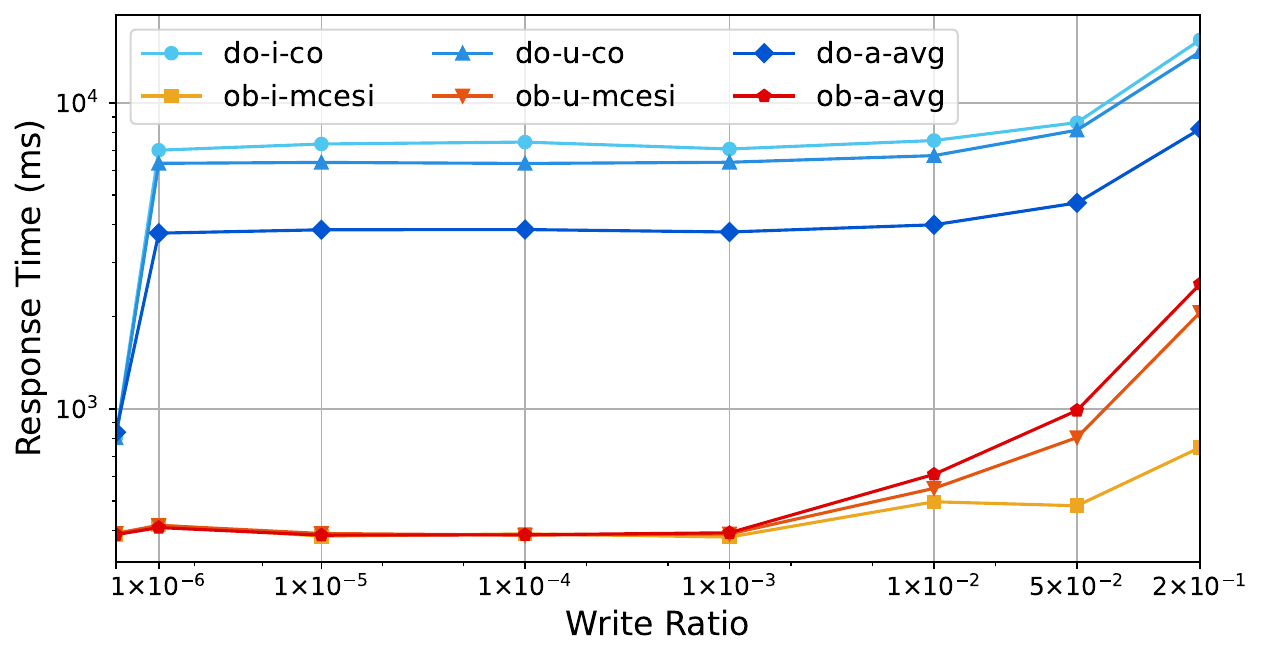}
    \vspace{-8pt}
    \caption{OceanBase vs. Doris under update-intensive workloads.}
    \label{fig:update_intensive_avg}
    \vspace{-15pt}
\end{figure}

We compared OceanBase Mercury with Doris based on practical update-intensive workloads. As shown in Figure~\ref{fig:update_intensive_avg}, the experimental results demonstrate the performance comparison between OceanBase and Doris under varying write-intensive workloads, measuring average response times across 18 queries for different write ratios ranging from 0.0 to 0.2. OceanBase consistently outperforms Doris, with an average response time of approximately 853ms compared to Doris's 4327ms, representing a roughly 5$\times$ performance advantage. As the write ratio increases from 0.0 to 0.2, the average response time escalates significantly from 613ms to 5388ms, indicating that write-intensive operations substantially degrade query performance for both systems. The figure visualizes these trends across different combinations of write types (insert/update) and table types (column, mow\_c\_encoding\_skip\_index), with response times displayed on a logarithmic scale to accommodate the wide performance range observed in the experiments.

\section{Related Work}
\label{sec:related}

\subsection{Near Real-time OLAP}

Near Real-time OLAP has gained significant attention for big data applications, including serving workloads such as monitoring and dashboard creation with complex analytical queries on newly added data. 
Community and commercial users of OceanBase frequently face such needs and traditionally have to involve multiple systems to meet the requirements.
To mitigate these issues, OceanBase Mercury (i.e., OceanBase 4.3.3) enhances its near real-time OLAP capabilities, functioning as a high-performance, unified, and flexible OLAP database. OceanBase Mercury aims to process analytical workloads within a single framework, enhancing efficiency and reducing resource wastage. The data to be processed can originate from OceanBase's OLTP system as well as various other sources, including other databases, Kafka, Flink, file systems, and more. Innovations in its column store mode, materialized view, and vectorized execution engine significantly improve OceanBase's analytical performance. With OceanBase version 4.3.3, users can handle large-scale analytical queries and serving tasks within a single system. This provides developers the capability to easily build a large-scale and near real-time OLAP system while supporting the OLTP capability.




OceanBase Mercury excels in in-memory processing, linear scaling, and high data compression, though it faces challenges with memory dependency and hybrid workload tuning. Compared with other open-source OLAP databases, Presto~\cite{PrestoICDE19} offers flexible multi-source integration but lacks UDF (User-Defined Function) support and struggles with memory demands. ClickHouse~\cite{schulze2024clickhouse} uses columnar storage and scalability for efficient analytics but has limited cluster expansion capabilities and high maintenance costs. Druid~\cite{DruidSIGMOD14} specializes in near real-time ingestion and multi-tenancy but suffers from inconsistent OLAP performance and complexity. StarRocks~\cite{StarRocks} combines vectorized execution and MPP architecture for sub-second queries but is constrained by storage-compute coupling and limited UDF or UDTF (User-Defined Table Function) support, reflecting trade-offs between performance, scalability, and ecosystem maturity across systems. 

Furthermore, Napa~\cite{agiwal2021napa} at Google powers scalable data warehousing by aggressively using materialized views maintained via LSM-tree structures. 
Napa introduces a \textit{Queryable Timestamp} (QT) to manage the trade-off between data freshness, query performance, and resource cost. 
To guarantee robust query latency (low variance), newly ingested data in Napa becomes visible only after the QT advances, which may be delayed until sufficient background compaction or view maintenance is complete. 
In contrast, Mercury achieves a freshness of zero to milliseconds (A2 system in \cite{zhang2024hybench}) by merging the columnar baseline with row-based incremental data during query execution. This design enables immediate access to committed transactions without waiting for compaction, prioritizing OLTP stability while maintaining analytical freshness.

\subsection{Column store}
Column store is very suitable for analytical tasks due to its unique structure and access mode. C-store~\cite{stonebraker2018c} is a read-optimized rather than write-optimized relational DBMS, and provides a projection method in different columns. HBase~\cite{vora2011hadoop} is a NoSQL distributed database system built on top of HDFS, and HDFS contains the non-textual data such as images while HBase stores the location of such data. ClickHouse~\cite{ClickHouse} is a high-performance, MPP-architecture, column-storage, and OLAP database, which uses all available system resources to their full potential to process each analytical query as fast as possible. In addition to the native column store databases mentioned above, there are also some column store-compatible databases. For example, Alibaba's Hologres~\cite{DBLP:journals/pvldb/JiangHXJJXJYWJM20} implements a row-column hybrid storage architecture, in which column store is used to optimize the performance of column scans. SQL Server supports column store index and gives an overview of the design and implementation of column store indexes in~\cite{larson2011sql}. And in~\cite{abadi2008column}, authors draw a conclusion that a row-store cannot obtain the performance benefits of a column-store even if we optimize a row-store by vertically partitioning the schema or by indexing every column so that columns can be accessed independently. The technology evolution and history of the fall of row stores and the rise of column stores is introduced in ~\cite{sridhar2017modern} by delving into architectural details of column DBs from academia and industry.

\subsection{Vectorized execution engine}

The vectorized execution engine has proven to be a powerful mechanism for significantly enhancing the performance of analytical queries, primarily due to its ability to process larger amounts of data with a single instruction. The concept of vectorized execution is first proposed in~\cite{boncz2005monetdb}, in which the Volcano-like tuple-at-a-time execution architecture of relational systems is recognized as the reason for low CPU efficiencies on modern CPUs because of the interpretation overhead. To address this issue, the MonetDB/X100 query engine was proposed in the same paper, which implemented vectorized execution on top of MonetDB, marking a pivotal advancement in query processing techniques. Following this breakthrough, vectorized execution gained widespread adoption across numerous database systems due to its remarkable performance benefits. For instance, CockroachDB~\cite{taft2020cockroachdb} incorporates both the traditional row-at-a-time execution engine and the vectorized execution engine, dynamically selecting the optimal engine based on the characteristics of the input data during query execution. Further advancements in this area are demonstrated in~\cite{shen2021using}, where a comprehensive vectorized execution engine is introduced for SQL query processing on Spark. This engine optimizes CPU cache utilization and minimizes data footprint, further improving efficiency. Additionally, a detailed comparison between vectorized execution and data-centric compiled query execution is presented in~\cite{kersten2018everything}, which evaluates their respective strengths and weaknesses across various metrics such as computation cost.

\subsection{Materialized view}
Unlike normal views, the materialized views store the results of some time-consuming queries. When performing relevant queries, the results can be directly accessed without having to repeat these time-consuming operations. Because of the convenience brought by materialized views, a lot of database systems support this feature. The Oracle Materialized Views~\cite{bello1998materialized} are designed for data warehousing and replication, in which the optimization with materialized views includes a transparent query rewrite based on a cost-based selection method. IBM's DB2 proposed an advanced tool to automatically select optimal materialized views and indexes to decrease the maintenance cost in 2004~\cite{zilio2004recommending}. An algorithm~\cite{quoc2016synchronous} in PostgreSQL is proposed to incrementally update the materialized views as the data in the base tables are changed. More specifically, the algorithm devised a program that automatically generates codes in PL/pgSQL for triggers, which can undertake synchronous incremental updates of the materialized views in PostgreSQL. Three crucial problems in the field of materialized views are discussed in ~\cite{chirkova2012materialized}, and the three problems are as follows: how to efficiently maintain materialized views when base tables change; how to efficiently improve query performance through materialized views; how to choose appropriate views for materialization.

\subsection{Update Models}

Table~\ref{tab:update-models} summarizes four prevalent update models employed in analytical processing systems, each representing distinct trade-offs between write and read performance. \textbf{Copy-on-Write} achieves optimal read performance by maintaining a single, consistent data version, but incurs high write costs as it requires reading and rewriting entire rowsets for each update. This model excels in read-heavy workloads where data freshness is critical, as exemplified by systems like Delta Lake~\cite{DBLP:journals/pvldb/ArmbrustDPXZ0YM20} and Snowflake~\cite{DBLP:conf/sigmod/DagevilleCZAA16}. \textbf{Merge on Read} minimizes write overhead by appending updates without immediate consolidation, deferring merge operations to query time. While this approach enables efficient writes, it suffers from degraded read performance when incremental data accumulates, requiring complex multi-version merging and deduplication during queries. Systems such as OceanBase~\cite{DBLP:journals/pvldb/YangXGYWZKLWX23}, TiFlash~\cite{huang2020tidb}, and ClickHouse~\cite{schulze2024clickhouse} adopt this model, prioritizing write throughput over read latency. Efficient distributed transaction processing~\cite{zhao2025efficient} and deadlock detection mechanisms~\cite{yang2023lcl, yang2025lcl+} are crucial for maintaining consistency in such systems. \textbf{Delta Store} employs an indexed delta structure to track updates, balancing write and read costs at medium levels. However, predicate pushdown becomes complex, and query optimization deteriorates when updates are frequent. \textbf{Merge-on-Write (delete+insert)} represents a hybrid approach: it marks deletions using delete-bitmaps while appending new data to separate rowsets, achieving low read costs through efficient filtering while maintaining moderate write costs. This model supports predicate pushdown and provides good query optimization, making it suitable for systems requiring both update efficiency and query performance, as demonstrated by Polar IMCI~\cite{DBLP:conf/fast/CaoLCZLWOWWKLZZ20}, SQL Server CSI~\cite{larson2011sql}, and Hologres~\cite{jiang2020alibaba}.

\begin{table*}[htp]
\centering
\caption{Update Models in AP Systems}
\label{tab:update-models}
\scriptsize
\begin{tabular}{lcp{2.4cm}cp{2.6cm}ccp{3.6cm}}
\toprule
\textbf{Model} & \textbf{Write} & \textbf{Write Process} & \textbf{Read} & \textbf{Read Process} & \textbf{Pred.} & \textbf{Query} & \textbf{Systems} \\
 & \textbf{Cost} & & \textbf{Cost} & & \textbf{Push.} & \textbf{Opt.} & \\
\midrule
Copy-on-Write & High & Read base, modify all, write new rowset & None & None & Yes & Good & Hudi CoW~\cite{Hudi}, Delta Lake, Snowflake \\
Merge on Read & None & None & High & Read multi-version rows, merge \& deduplicate & Complex & Poor (large incr.) & Hudi MoR~\cite{Hudi}, StarRocks Unique, OceanBase, TiFlash, ClickHouse \\
Delta Store & Med. & Find row via index, update delta & Med. & Merge base \& delta by row number & Complex & Poor (freq. updates) & Apache Kudu~\cite{lipcon2015kudu} \\
Merge-on-Write & Med. & Find row via index, mark delete, write new & Low & Filter deleted via delete-bitmap & Yes & Good & SQL Server CSI, ADB, Polar IMCI, Hologres, StarRocks PK \\
\bottomrule
\end{tabular}
\end{table*}

\section{Lessons and Future Innovations}
\label{sec:lesson}

We compile the lessons learned from developing OceanBase Mercury to inform future OLAP innovations.

\textbf{Lesson 1.} Simultaneously accommodating multiple storage models offers users more options, but also poses heightened challenges for the optimizer. This necessitates a more intricate comprehension and assessment of the diverse operator costs within different models to select more effective plans.

\textbf{Lesson 2.} An overabundance of incremental data can have a substantial impact on query response time, whereas frequent compactions generate substantial drain on computing resources and adversely affect user operations. Enhanced monitoring and feedback mechanisms are necessary to optimize query performance through compaction without disrupting business operations.

\textbf{Lesson 3.} The refresh of full or incremental materialized views typically involves handling large volumes of data. Therefore, it is essential to utilize parallel DML and rapid bulk data loading techniques to enhance the refresh speed.

\textbf{Lesson 4.}	Each incremental refresh of a materialized view entails a substantial number of row deletions. It is crucial to employ an efficient method, such as TTL (Time-to-Live), to manage these deletions effectively.

\textbf{Lesson 5.}	To further accelerate calculations, the computing engine should consider expressions and aggregations, joins, sorts, and other operations based on encoded data (such as dictionary data).

\textbf{Lesson 6.}	ARM chips offer higher overall cost performance, and further consideration needs to be given to optimizing the computing engine in the ARM environment.

\section{Conclusions}
\label{sec:conclusion}

In conclusion, the rapidly evolving landscape of big data necessitates database systems that can seamlessly handle vast volumes of data while providing transactional processing, near real-time serving, and robust analytical capabilities. Traditional OLAP systems often fall short in addressing these demands, leading to the development and adoption of near real-time analytical processing systems. This paper introduces the OceanBase Mercury database system, designed to operate at petabyte scale with near real-time OLAP capabilities. OceanBase Mercury addresses the challenges of maintaining ACID properties across distributed transactions, enabling complex analytical queries, and providing high availability, scalability, and robust performance. The proposed system offers a flexible hybrid column storage mode, efficient refresh mechanisms for materialized views, three data formats for the vectorized engine, and superior performance under various workloads, demonstrating the advantages of OceanBase Mercury in addressing the evolving needs of enterprises for efficient data processing and analysis.

\section*{Acknowledgment}

We would like to thank the anonymous reviewers for their insightful comments, and our colleagues for their contributions. The source code is available at https://github.com/oceanbase/oceanbase/tree/4.3.3.



\bibliographystyle{IEEEtran}
\bibliography{IEEEabrv, IEEEexample}

\end{document}